\documentclass[article,submit,moreauthors,pdftex]{mdpi} 

\firstpage{1} 
\pubvolume{xx}
\issuenum{xx}
\articlenumber{xx}
\pubyear{2021}
\copyrightyear{2021}

\pdfoutput=1

\usepackage[caption=false]{subfig} 
\usepackage{fancyhdr}
\usepackage[yyyymmdd,hhmmss]{datetime}

\Title{Assimilating X- and S-band Radar Data for a Heavy Precipitation Event in Italy}

\Author{Valerio Capecchi$^{1,*}$, Andrea Antonini$^{1}$, Riccardo Benedetti$^{1}$, Luca Fibbi$^{1,2}$, Samantha Melani$^{1,2}$, Luca Rovai$^{1,2}$, Antonio Ricchi$^{3}$ and Diego Cerrai $^{4,5}$}

\AuthorNames{Valerio Capecchi, Andrea Antonini, Riccardo Benedetti, Luca Fibbi, Samantha Melani, Luca Rovai, Antonio Ricchi and Diego Cerrai}

\address{%
$^{1}$ \quad LaMMA - Laboratorio di Meteorologia e Modellistica Ambientale per lo sviluppo sostenibile, Italia; capecchi@lamma.toscana.it\\
$^{2}$ \quad Istituto per la BioEconomia - Consiglio Nazionale delle Ricerche, Italia\\
$^{3}$ \quad Department of Physical and Chemical Sciences, Universit\'a degli Studi dell'Aquila, Italia\\
$^{4}$ \quad Department of Civil and Environmental Engineering, School of Engineering, University of Connecticut, USA\\
$^{5}$ \quad Eversource Energy Center, University of Connecticut, USA}

\corres{Correspondence: capecchi@lamma.toscana.it; Tel.: +39-055-448-301}

\abstract{During the night between 9 and 10 September 2017, multiple flash floods associated to a heavy-precipitation event affected the town of Livorno, located in Tuscany, Italy. Accumulated precipitation exceeding 200 mm in two hours, associated with a return period higher than 200 years, caused all the largest streams of the Livorno municipality to flood several areas of the town. We used the limited-area Weather Research and Forecasting (WRF) model, in a convection-permitting setup, to reconstruct the extreme event leading to the flash floods. We evaluated possible forecasting improvements emerging from the assimilation of local ground stations and X- and S-band radar data into the WRF, using the configuration operational at the meteorological center of Tuscany region (LaMMA) at the time of the event. Simulations were verified against weather station observations, through an innovative method aimed at disentangling the positioning and intensity errors of precipitation forecasts. 
By providing more accurate descriptions of the low-level flow and a better assessment of the atmospheric water vapour, the results demonstrate that assimilating radar data improved the quantitative precipitation forecasts.}

\keyword{WRF model; 3D-Var data assimilation; radar data; short-range prediction; heavy precipitation event
 }

\begin{document}

%
\section{Introduction}
Almost every year, during the fall season, heavy precipitation events (HPEs) bring destruction and cause fatalities somewhere in the Western Mediterranean (WM) region \citep{gaume2009compilation,llasat2010high,llasat2013towards,insuacosta2021climatology}. HPEs can either persist for several days on large areas, resulting in extensive flooding (e.g. 1966 Arno, Italy \cite{malguzzi20061966,capecchi2019reforecasting} and 1994 Piedmont, Italy \cite{buzzi1998numerical,capecchi2020reforecasting}), or manifest at the sub-daily scale and produce flash floods (e.g. 1999 Aude, France \citep{nuissier2008numerical,ducrocq2008numerical} and 2011 Liguria, Italy \cite{buzzi2014heavy,capecchi2021reforecasting}).

Short duration HPEs often assume the form of V-shaped, quasi-stationary convective systems \cite{fiori2014analysis,fiori2017triggering}. These storms are characterized by a back-building dynamics, where convective cells developing upstream of the affected area continuously replace dissipating cells, resulting in pulsating heavy rain \cite{chappell1986quasi,doswell1996flash}. This continuous cell replacement is frequently initiated offshore \cite{duffourg2016offshore}. This structure becomes quasi-stationary due to the perduration of a favorable synoptic and mesoscale environment \cite{romero2000mesoscale}. Important factors for persistence and evolution of these precipitation phenomena can be found at the surface (e.g. moist and conditionally unstable air associated to warm sea surface temperatures \cite{cassola2016role,ducrocq2014hymex}), within the boundary layer (e.g. convergence associated to a frontal system \cite{weaver1979storm}, induced by orography \cite{fiori2014analysis,buzzi2014heavy,ducrocq2014hymex} or land-sea differences \cite{cassola2016role,davolio2009high}), and throughout the troposphere (e.g. coupling between low and upper level jet streaks \cite{uccellini1979coupling}). When these conditions persist for several hours, the system may assume the characteristics of a mesoscale convective system, which is able to partially drive the mesoscale circulation \cite{zipser1982use,fritsch2001mesoscale,houze2004mesoscale}.

The frequency of these HPEs over the WM region has already increased \cite{alpert2002paradoxical,reale2013synoptic}, although a significant reduction of annual rainfall has been observed in relatively recent years \cite{piervitali1998rainfall,brunetti2004changes}. Climate projections support this trend, by predicting further increase in frequency of HPEs, and decrease of the average precipitation in the WM region \cite{scoccimarro2016heavy}. In this framework, considering that in the period 1975-2001 the average mortality per flash floods was 5.6\% of the affected population \cite{jonkman2005global}, the most threatening consequence of the increasing HPE frequency is the expected increase of human losses associated to the flash floods they produce. Monitoring, mitigating, predicting and communicating the potential of such dramatic events is necessary for reducing their mortality \cite{doocy2013human}.

Several studies investigated improvements in the characterization of HPEs precipitation fields arising from Three-Dimensional Variational (3D-Var) assimilation of radar and ground observations into the Weather Research and Forecasting (WRF) model (\cite{maiello2014impact,sugimoto2009examination,lagasio2019predictive,xiao2007multiple,tian2017assimilation}). The vast majority of these studies used weather analysis as initial and boundary conditions and assimilated data which were measured during the events. This methodology, despite being optimal for an accurate reconstruction of HPEs, is not suitable for evaluating operational improvements. For bridging the gap between the improvements described in previous works and operational improvements, a reasonable latency consisting of observations collection, data assimilation, WRF run time, and forecast communication for early warning should be taken into account. Considering this latency by assimilating observations before - and not during - the storm allows for a quantification of the impact data assimilation has on the predictability of HPEs in a pseudo-operational framework.

This work focuses on the numerical reconstruction of the quasi-stationary convective system that, during the night between 9 and 10 September 2017, caused several flash floods in the town of Livorno, Italy, resulting in nine fatalities. The Livorno case was recently studied by \cite{federico2019impact} using the non-hydrostatic model RAMS@ISAC \cite{cotton2003rams}. The authors evaluated the impact of assimilating lightning and radar reflectivity  data in the very short-term (i.e., forecast length shorter than 3 hours) and at the convection-permitting scale (i.e., grid spacing up to about 1 km). The authors underline the paramount importance of the reflectivity data and found that the assimilated runs outperformed the control one (i.e., no data assimilated), by reducing the number of missed precipitation events at the expense of a higher number of false alarms. The authors acknowledged that not updating the initial and boundary conditions in the assimilation step represents a limit of their work, and they claimed that exploring this issue deserves further investigations. \cite{lagasio2019synergistic} tested the effect of assimilating a broad range of remote sensed data into the WRF model. The authors found that information about the wind field and, to a minor extent, atmospheric water vapour content are crucial for a better localization of the rainfall peaks.

We present the simulation of the Livorno case using the WRF model in an operational-like configuration. We investigate the impact of assimilating radar and automatic weather station data for a refinement of the operational setup, which will lead to immediate benefits for any early warning system that relies on data produced by numerical weather  models. In particular, the main goal of the work is to evaluate to what extent the information carried by a relatively small radar system, such as the X-band radar located at the Livorno harbor \cite{Antonini2017}, is valuable to improve the predictions of  the WRF model in the short-term. In fact, in the recent years some European projects, such as those funded in the framework of the European Cross-Border Cooperation Programme Italy-France ``Maritime'', dealt with the deployment of a network of X-band radars in the Tuscany region and surrounding areas. The explicit goal of such projects is to monitor potential severe weather, occurring offshore, using radar instruments as an essential tool for nowcasting applications. The cross-border sharing of such relevant meteorological observations and the integration  with existing tools and methodologies is intended to improve the forecasting and alerting capabilities of operational regional weather agencies.

The paper is organized as follows: in Section \ref{sec:matmet} we describe the meteorological context of the Livorno case and the modelling setup implemented, in Section \ref{sec:pacman} we give the details about the method used to evaluate model outputs, in Section \ref{sec:results} we present the main results, and in Section \ref{sec:discuss} we discuss  implications of the results, limitations of this study and we explain our vision of the path forward in the last Section.
%
\section{Materials and Methods}\label{sec:matmet}
%
\subsection{Synoptic conditions}
\begin{figure}[H]
\includegraphics[height=0.99\textwidth,angle=-90]{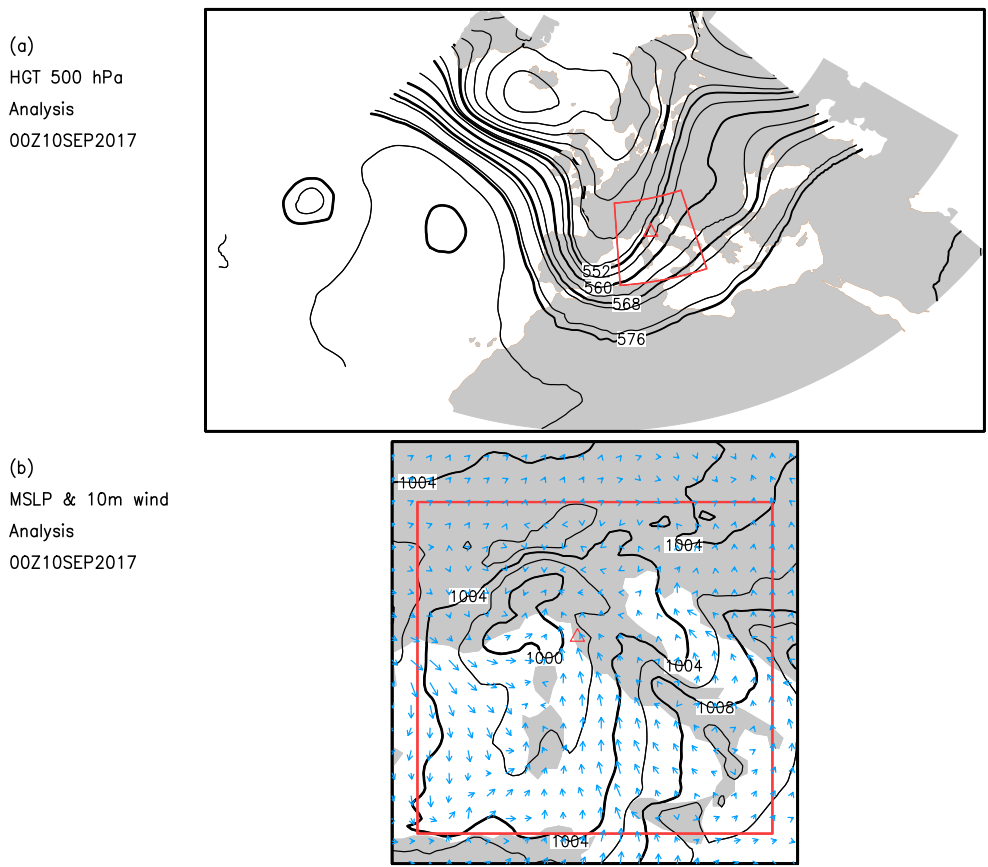}
\caption{(\textbf{a}): geopotential height at 500 hPa isobaric level in decametre at 0000 UTC of the 10th of September 2017. (\textbf{b}): mean sea level pressure in hPa and 10-metre wind vectors in blue color at the same time. In both panels, the red rectangle indicates the WRF domain of integration and the red triangle (``\textcolor{red}{$\bigtriangleup$}'') indicates the location of the  Livorno town. ERA5 data \cite{hersbach2020era5} were used to plot the maps.}\label{fig:syn}
\end{figure}

During the first hours of the 9th of September, a large trough deepening over the Eastern Atlantic Ocean approached the Mediterranean Sea. At 0000 UTC of the 10th of September 2017, the axis of the trough was oriented from the Scandinavian Peninsula to the Western Mediterranean Sea (see Figure \ref{fig:syn}a). The trough slowly moved eastward causing the deepening of a low pressure area (see Figure \ref{fig:syn}b) over the Ligurian sea and lee of the Italian Alps \citep{buzzi1978cyclogenesis,buzzi2020cyclogenesis}. As a result the pre-existing warm and humid air masses over the Tyrrhenian Sea interacted with the dry and relatively cold flow from France (see Figure \ref{fig:syn}b and Figure 7b in \cite{federico2019impact}), making the environment conducive for persistent precipitation systems. Furthermore, upper level divergence, sustained convective available potential energy values in the range 500-1000 J$\cdot$ kg$^{-1}$ and strong wind shear (plots not shown) favoured convective motions. A well-defined line of convergence of low-level winds over the Tyrrhenian Sea acted as a trigger to overcome the convective inhibition energy (see blue wind vectors in Figure \ref{fig:syn}b).
Following the classification of \cite{molini2011classifying}, the Livorno case belongs to the HPEs having a short duration (less than 12 hours) and a spatial extent smaller than $50 \times 50$ km$^2$. 
%
\subsection{Observed precipitation measurements}
\begin{figure}[H]%
 \centering
 \subfloat[]{\includegraphics[height=0.48\textwidth,angle=-90]{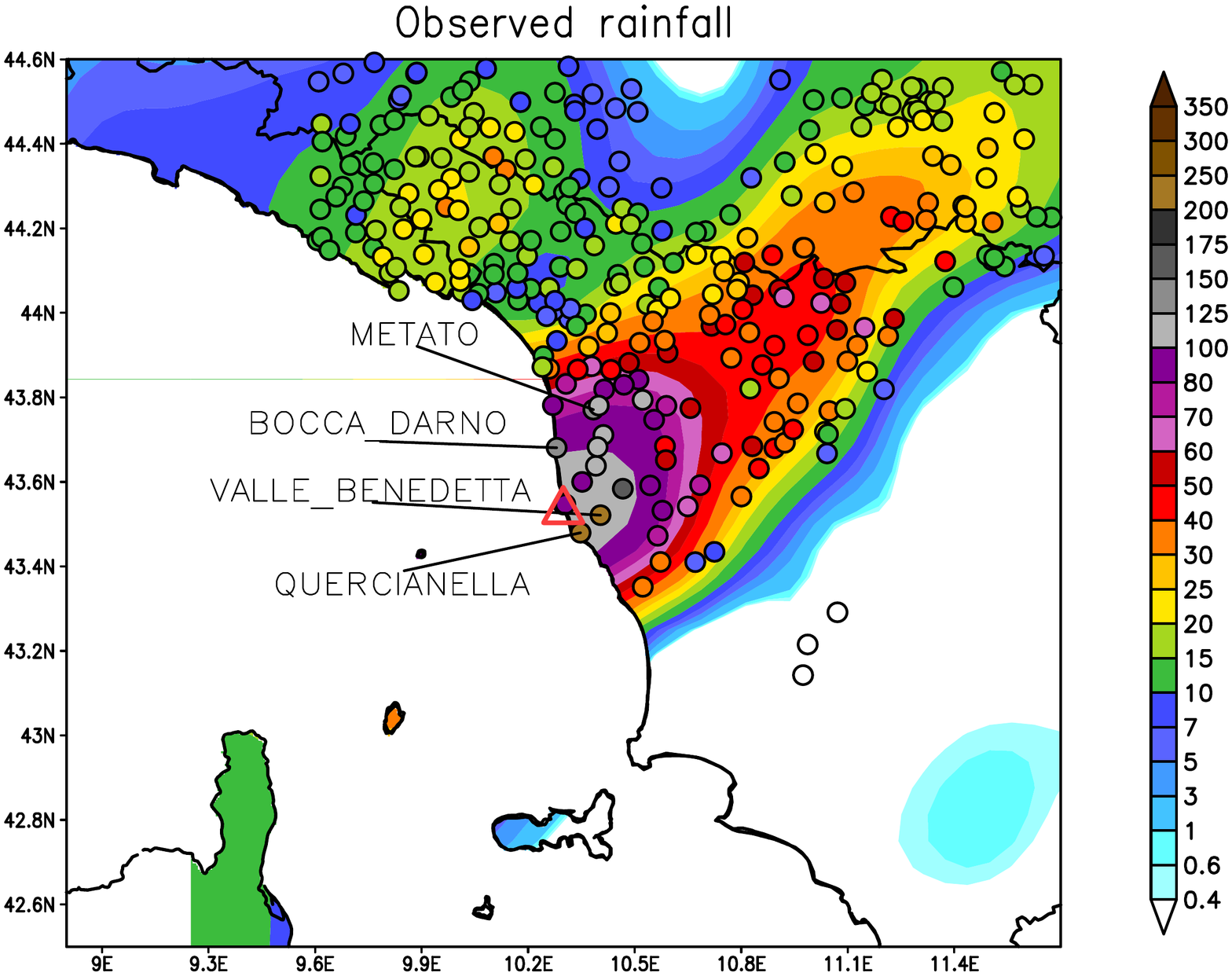}\label{fig:1a}}~~
 \subfloat[]{\includegraphics[height=0.48\textwidth,angle=-90]{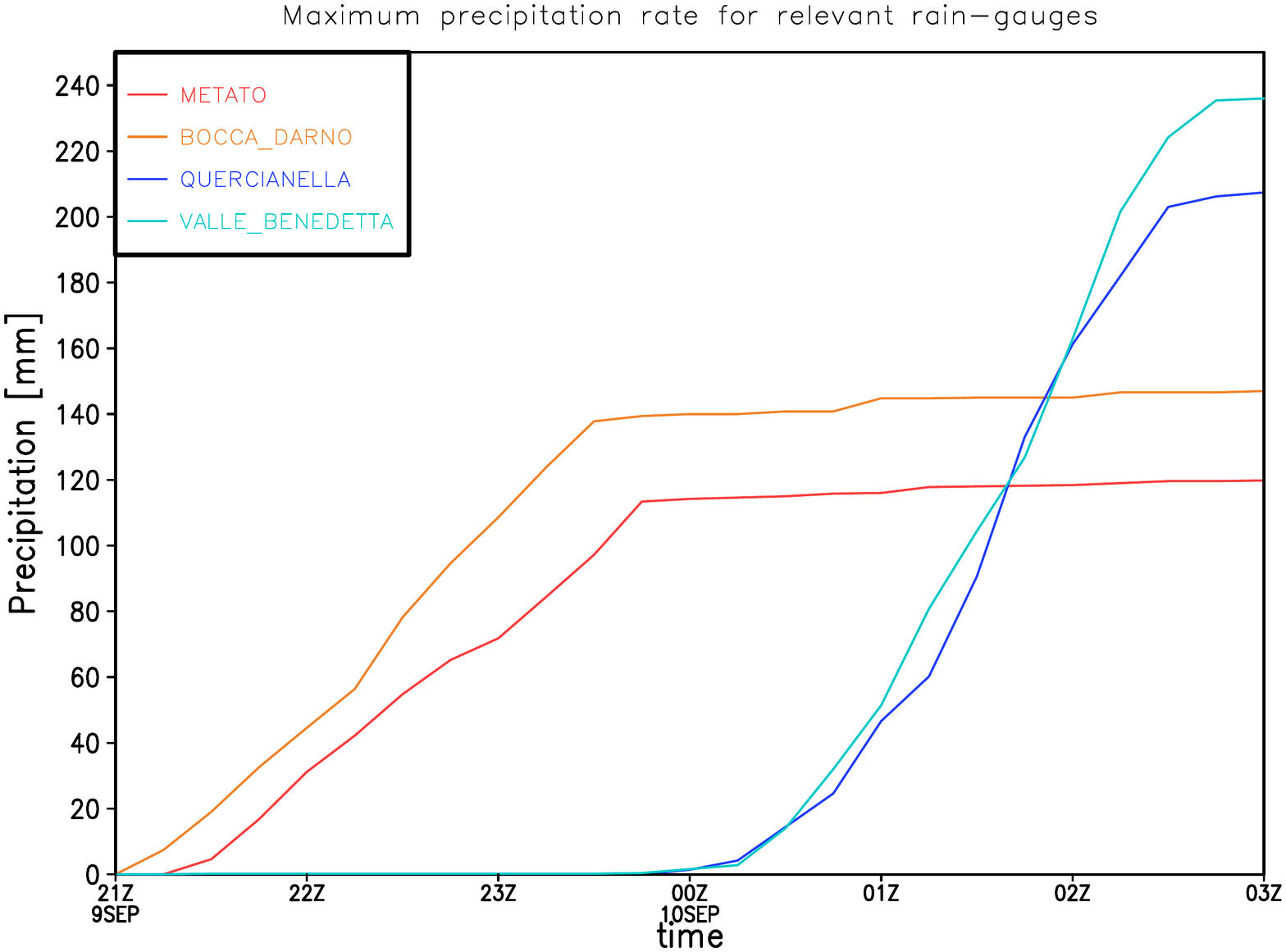}\label{fig:1b}}\\
 \caption{(\textbf{a}): observed rainfall accumulated in the 6-hour period ending on the 10th of September 2017 at 0300 UTC registered by automatic weather stations in the Tuscany region and surrounding areas. The Cressman interpolation technique \cite{cressman1959operational} is used to estimate rainfall amounts (shown with the shaded colours) when no rain-gauge data are available. The Livorno town is indicated with the red triangle  (``\textcolor{red}{$\bigtriangleup$}''). (\textbf{b}): accumulated precipitation registered by four rain-gauges located close to the Livorno town. Two of them are located to the North of the city (Metato and Bocca d'Arno) and two to the South (Valle Benedetta and Quercianella).}\label{figure_one}
\end{figure}

The Livorno case can be described by using the data obtained from the weather stations managed by the Hydrological Service of Tuscany (www.sir.toscana.it, link accessed in April 2021). The network is composed by approximately 400 rain-gauges, located over an area of about 23000 km$^2$, reporting conditions every 15 minutes \citep{capecchi2012fractal}. 
 In Figure \ref{figure_one}a we report the precipitation map of the accumulated precipitation in the 6-hour period ending on 0300 UTC of 10 September 2017. One precipitation maxima, located in the coastal area, and exceeding a cumulative rainfall amount of about 240 mm can be noticed, to the South of the Livorno town (indicated with the red triangle in Figure \ref{figure_one}a). The extreme is located in a widespread area of total precipitation exceeding 100 mm, which covers the entire central coast of Tuscany.
The variability of precipitation duration and intensity across the territory played a major role in the downstream hydrological responses. The sector South of the Livorno town, which was responsible for the flash floods in the urban areas, produced a precipitation that was limited in time: 210.2 mm were registered in the 2-hour period between 10 September, 0045 UTC and 0245 UTC (see Figure \ref{figure_one}b reporting the data collected at the station of Valle Benedetta). This corresponds to 86\% of the total event precipitation registered by this station. Another weather station, Quercianella, still located in the Southern sector, measured similar values: 188.6 mm in 2 hours, corresponding to 89\% of the total precipitation during the event. Analogies in the maximum hourly rain rate are even stronger, since the first station measured 120.8 mm in 1 hour, and the second 1 mm more. For both rain-gauges, \cite{regtos2017} estimated a return period $> 200$ years for the 1 and 3 hours duration rainfall events. The sector North of the town of Livorno, represented in Figure \ref{figure_one}b by the stations of Bocca d'Arno and Metato, was characterized by a peak rain rate between 50 and 70 mm in 1 hour and a total event duration of 3 to 4 hours. This precipitation was mostly concentrated between 2000/2100 UTC of the 9th of September and 0000 UTC of the 10th. In the rest of the region (plots not shown), precipitation was either intense and brief (further South: Castellina Marittima rain-gauge, return period $>200$ years for 1 and 3 hours duration rainfall events \cite{regtos2017}), or long lasting and less intense (North of Livorno: Coltano and Stagno rain-gauges, return period $\simeq90$ years for 12 hours duration rainfall events \cite{regtos2017}). 
%
\subsection{Radar data}
Several weather radars were operational during the rainfall event, belonging to different networks: the Italian national \cite{vulpiani2008italian,Vulpiani2012onthe}, the French national \cite{Tabary2007thenewfrench} and the Tuscany regional \cite{Antonini2017,cuccoli2020weather} ones. Among these available radars, two have been selected for this work: the Aleria and the Livorno radars, whose  locations are sketched in Figure \ref{fig:WRFdomain} with the red triangle. The former is a Doppler S-band (the signal frequency is 2802 MHz) system located in Aleria in the central Eastern Corsica and was detecting the severe weather system from a distance of approximately 70-80 km. The latter is a non-Doppler single polarization X-band (frequency is 9410 MHz) radar located in the Livorno harbor, and monitored the dynamics of the meteorological event moving towards the radar site. These systems have been designed and produced by different manufacturers and have  different technical characteristics, with particular reference to the operational frequencies, system dimensions and architectures \citep{Antonini2017}. The Aleria radar (S-band) is characterized by low attenuation impacts on radio signals due to the atmosphere and rainfall and consequently can reliably monitor rainfall fields up to a distance of approximately 100-150 km. The Livorno radar system (X-band) is more sensitive to smaller rainfall drops and can provide a finer resolution, but with a limited range due to the strong effects of atmosphere and rainfall attenuation on radio signals. Moreover the effects of path and wet radome attenuation due to the rain falling directly over the radar system are increased in this case by the strong precipitation occurring at the Livorno site. 
\begin{figure}[H]%
 \centering
 \subfloat[Livorno 2017/09/10 1730 UTC]{\includegraphics[height=0.3\textwidth]{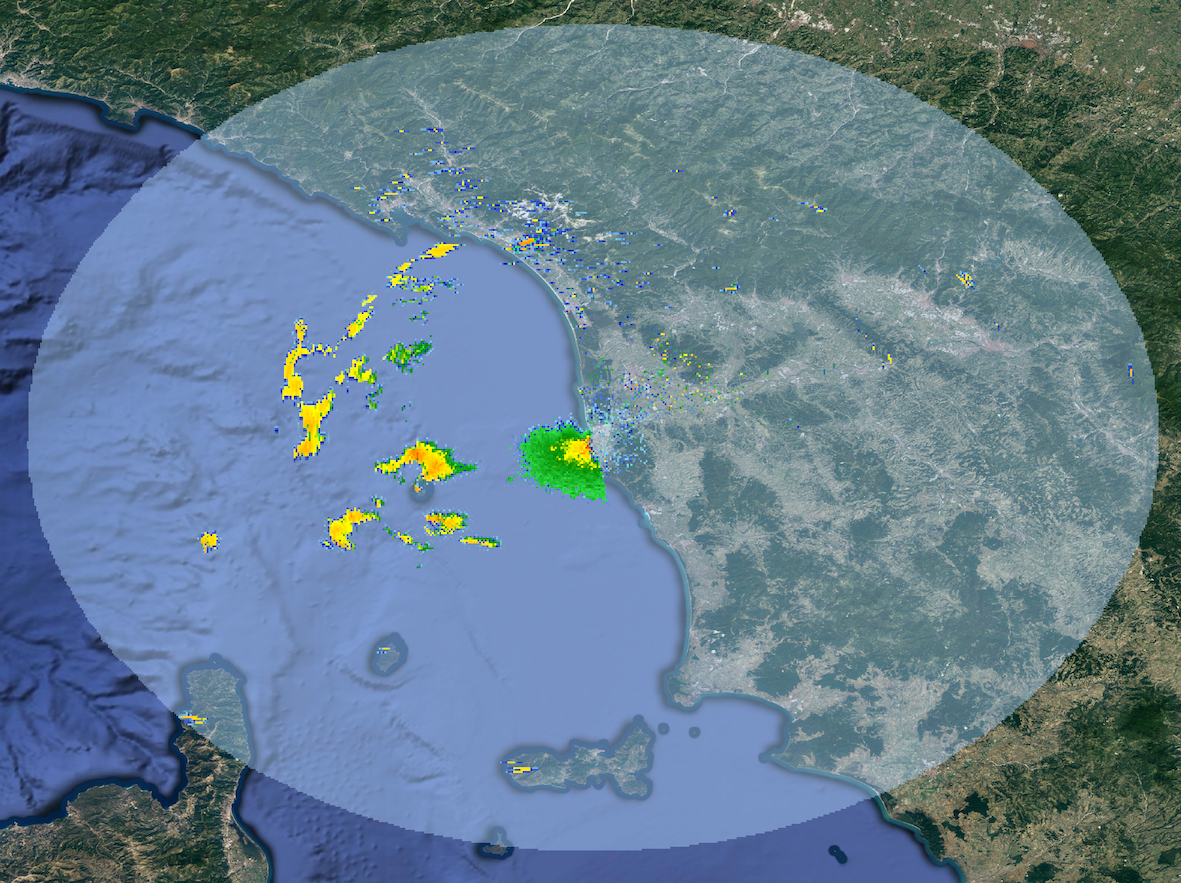}\label{fig:2a}}~~
 \subfloat[Aleria 2017/09/10 1730 UTC]{\includegraphics[height=0.3\textwidth]{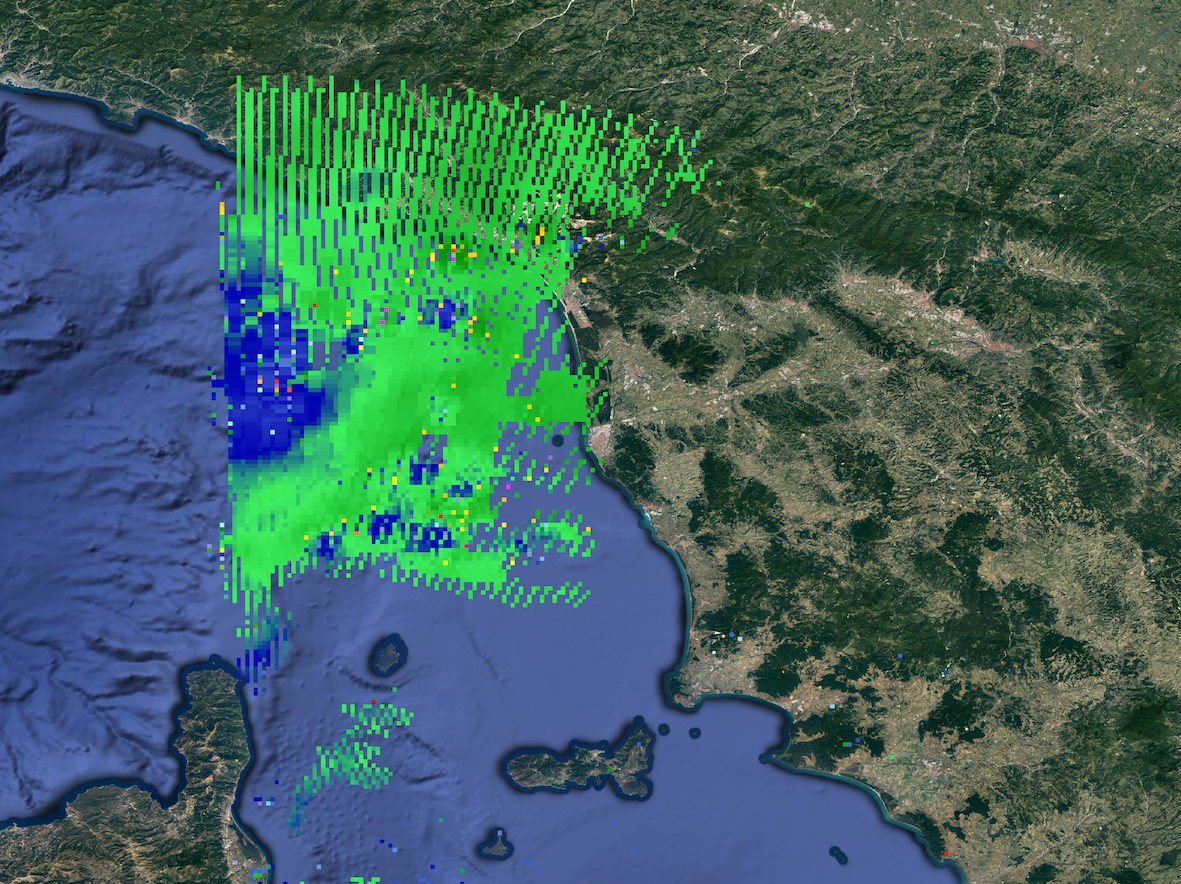}\label{fig:2b}}\\
 \subfloat[Livorno 2017/09/10 2045 UTC]{\includegraphics[height=0.3\textwidth]{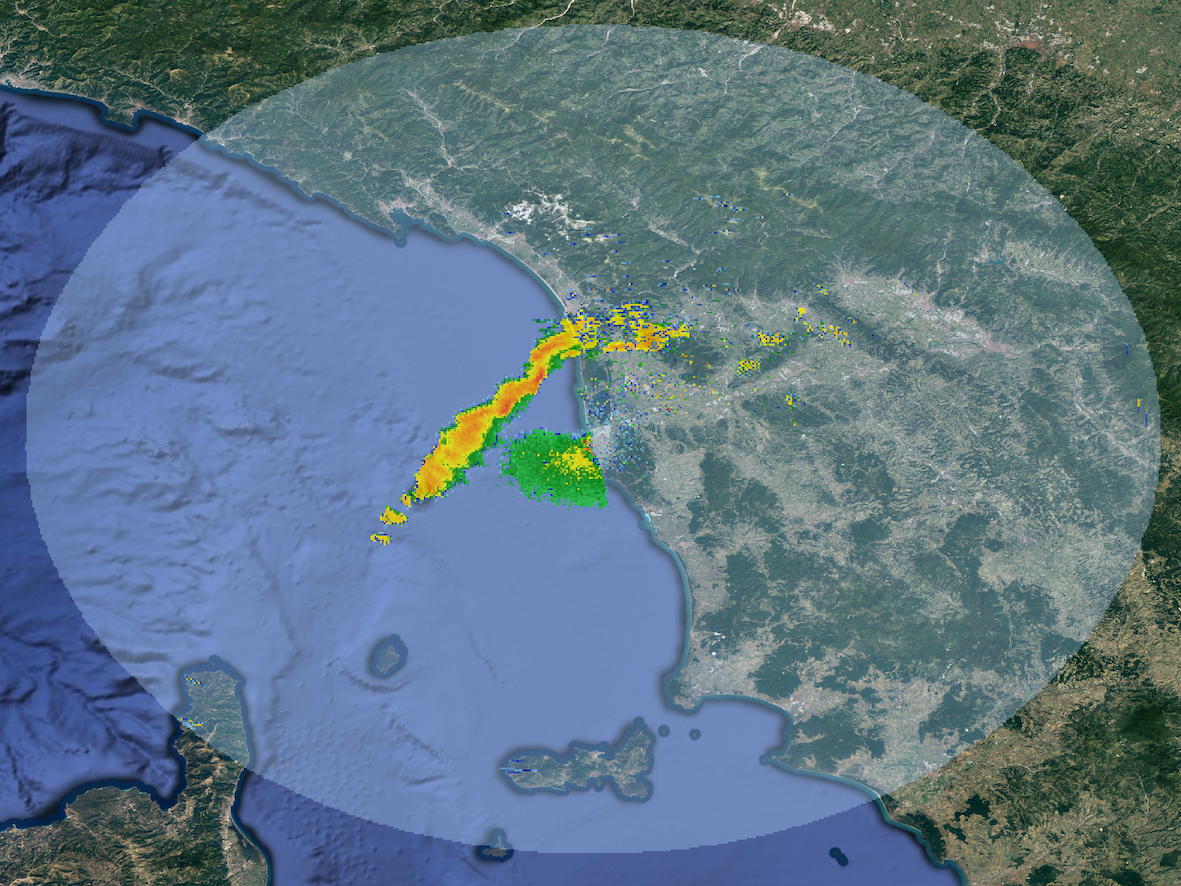}\label{fig:2c}}~~
 \subfloat[Aleria 2017/09/10 2045 UTC]{\includegraphics[height=0.3\textwidth]{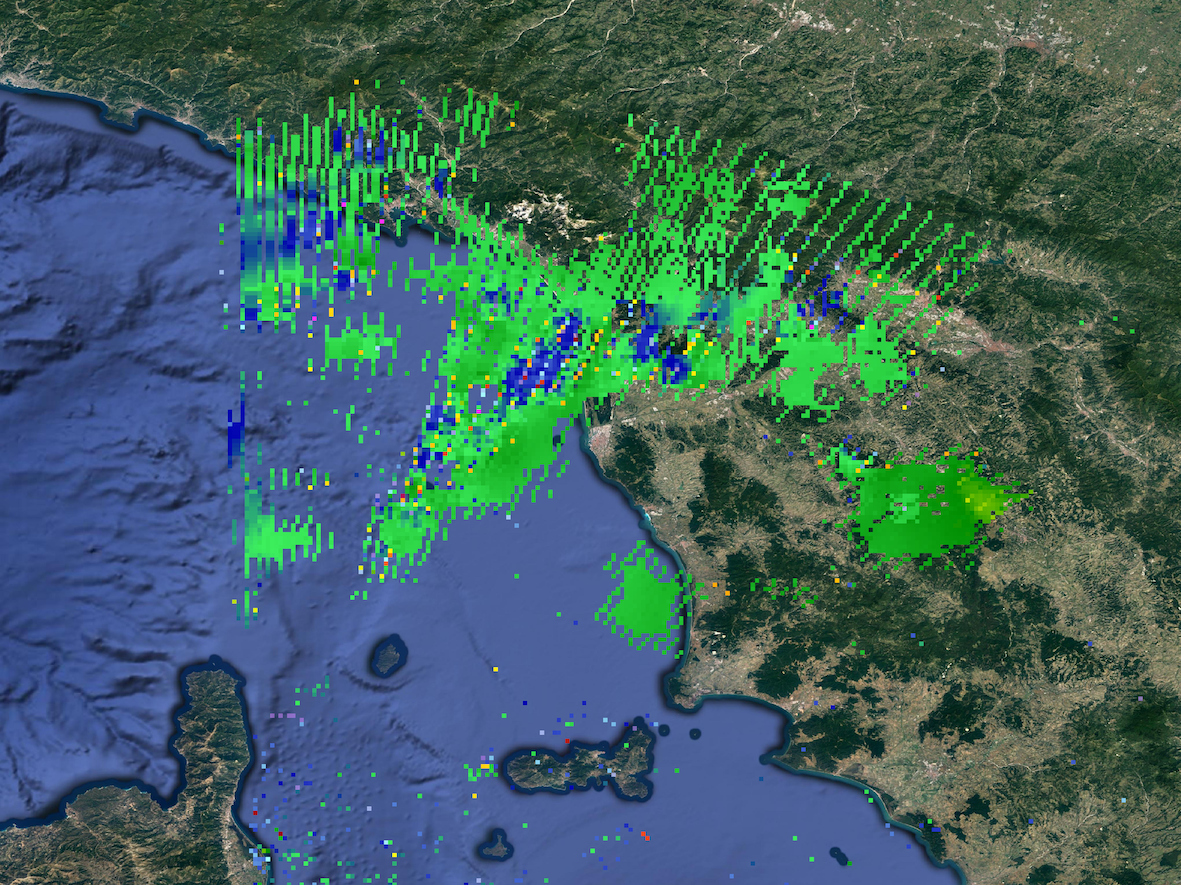}\label{fig:2d}}\\
 \subfloat[Livorno 2017/09/10 0100 UTC]{\includegraphics[height=0.3\textwidth]{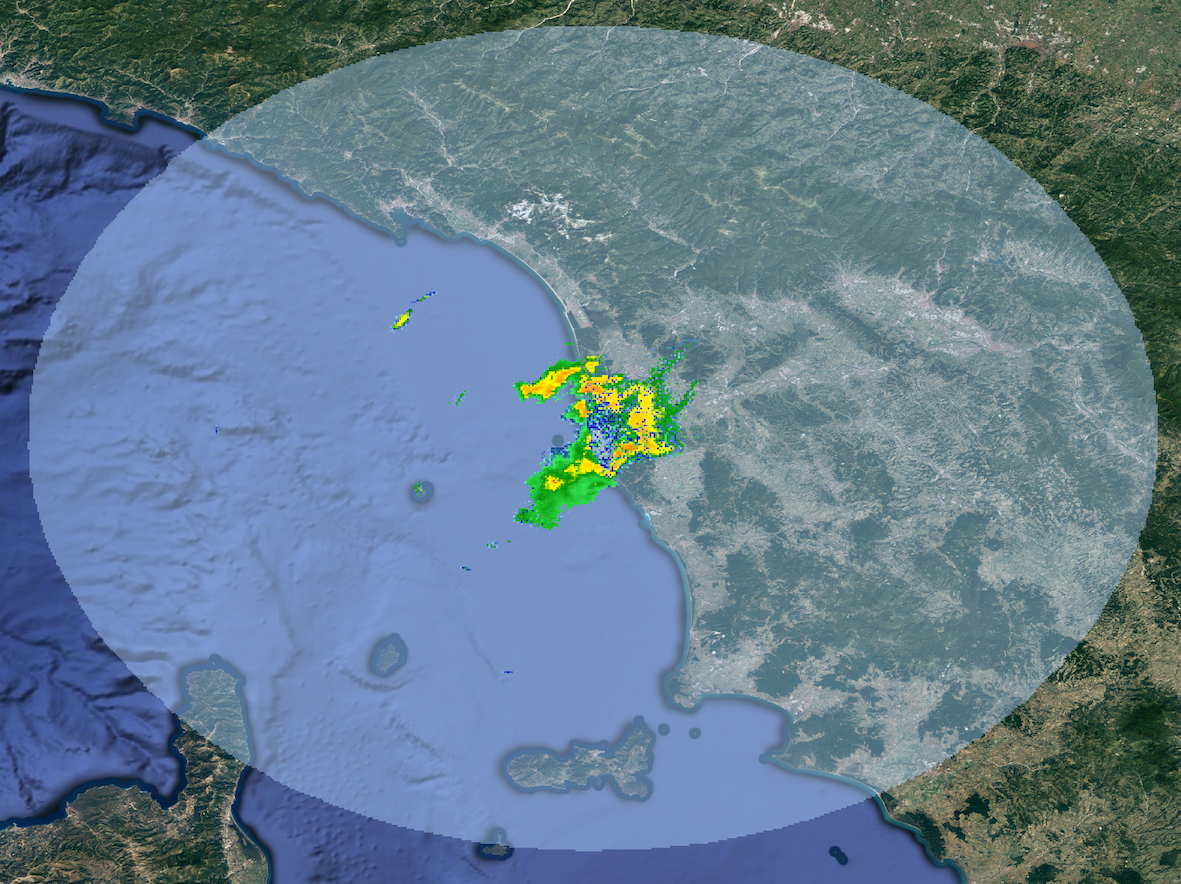}\label{fig:2e}}~~
 \subfloat[Aleria 2017/09/10 0100 UTC]{\includegraphics[height=0.3\textwidth]{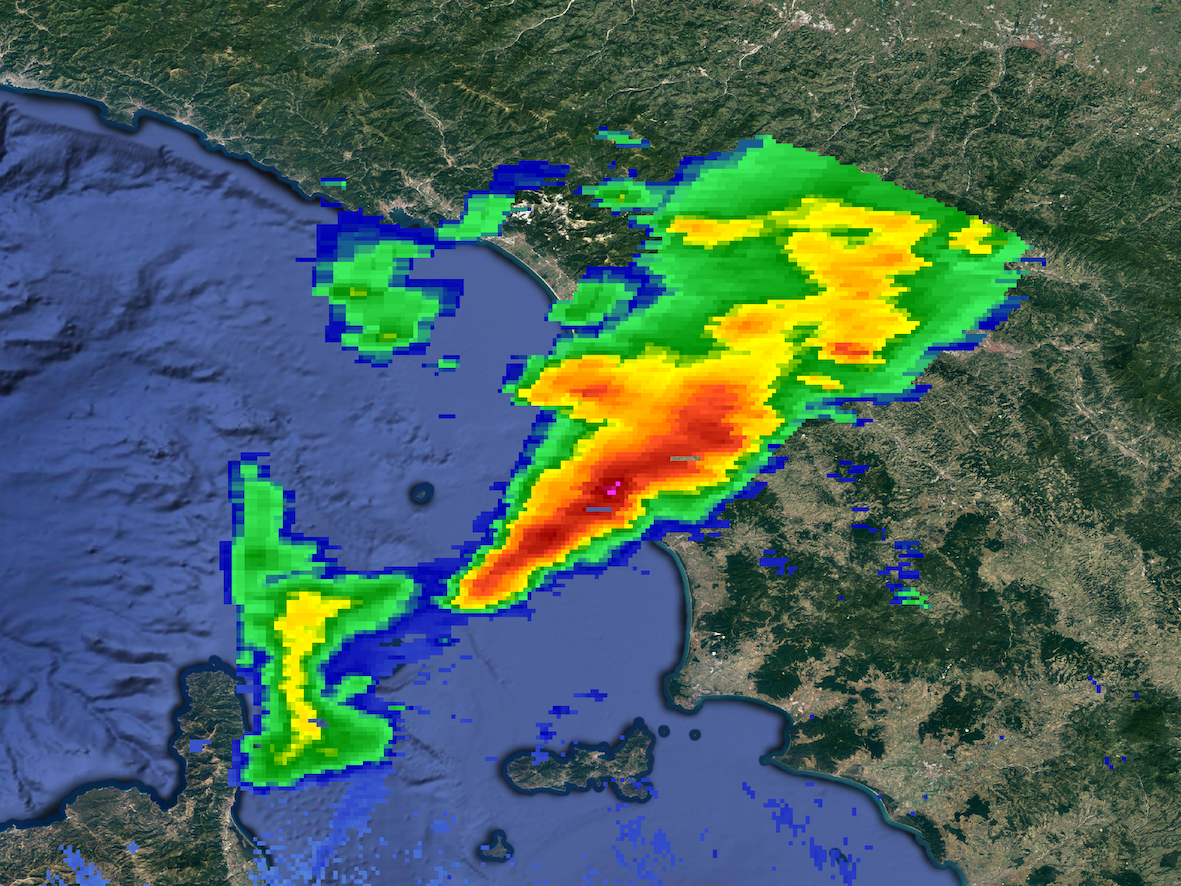}\label{fig:2f}}
 \caption{Sequence of radar reflectivity (unit dBZ) maps during the rainfall event as detected by the Livorno (left) and Aleria (right) radar systems. Panels \textbf{a} and \textbf{b} at 1730 UTC on the 9th of September, panels \textbf{c} and \textbf{d} at 2045 UTC on the 9th of September, panels \textbf{e} and \textbf{f} at 0100 UTC on the 10th of September.}
 \label{fig:radar}%
\end{figure}
These two radar systems detected the rainfall event since its origin over sea between North Corsica and Ligurian Gulf, while approaching the Tuscany coasts. The Livorno radar localized the weather event and followed its dynamics with high spatial detail, as shown in Figure \ref{fig:radar}a. Then the longitudinal shape of the strong precipitation front, well detected when approaching the Tuscany coasts (Figure \ref{fig:radar}c), underestimated the reflectivity intensity of the storm, at least during its beginning and early development on the sea (Figures \ref{fig:radar}b and d). Later, when the storm reached Livorno, the Aleria radar detected the strong rainfall occurrence and  persistence over the area (Figure \ref{fig:radar}f), while the Livorno radar system underestimated the intensity of precipitation, presumably due to a significant path and radome attenuation (Figure \ref{fig:radar}e). These two radars, in association with the Italian weather radar network \cite{vulpiani2008italian,Vulpiani2012onthe}, were used by  \cite{cuccoli2020weather} to characterize the Livorno event and were found to provide valuable information for the quantitative precipitation estimation.
%
\subsection{Satellite and lighting data}
The deep convection of the Livorno event was detected by the  Rapid Scan High Rate (5-minute) mode of the Spinning Enhanced Visible and Infrared Imager on board the Meteosat Second Generation satellite \cite{schmetz2002introduction}. Satellite images  (maps not shown) indicate a cloud top brightness temperature below -65$^\circ$ C, a temperature that can be found at approximately 12.5 km height, according to the atmospheric sounding measured at 0000 UTC on the 10th of September at Pratica di Mare (data not shown), a sounding station located approximately 250 km South of the area of interest. Further and in-depth analyses of the Livorno case using satellite-based observations are available in \cite{ricciardelli2018analysis}.

\begin{figure}[H]%
 \centering
 \subfloat[]{\includegraphics[width=0.5\textwidth]{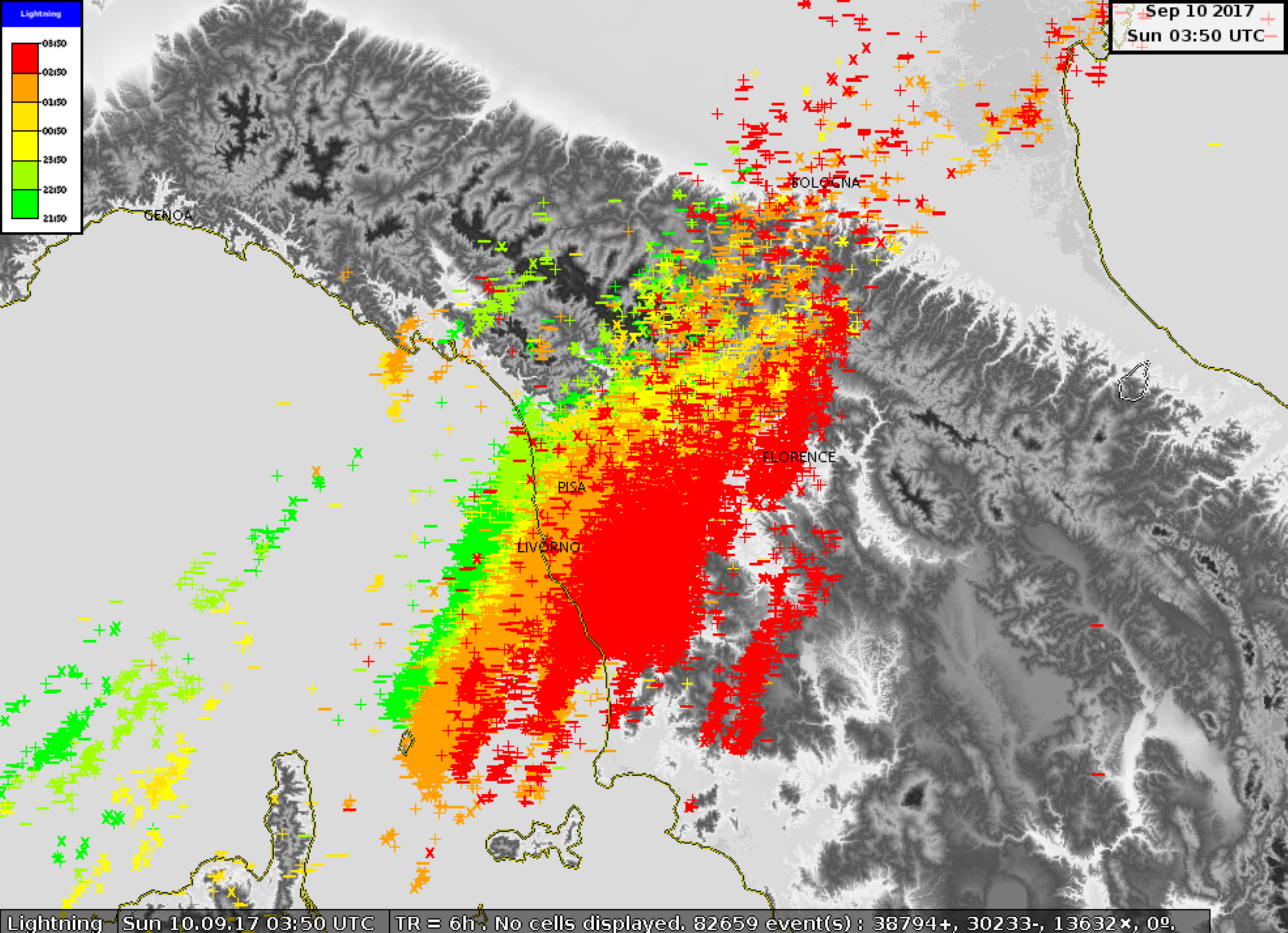}\label{fig:4b}}\\
 \subfloat[]{\includegraphics[width=0.5\textwidth]{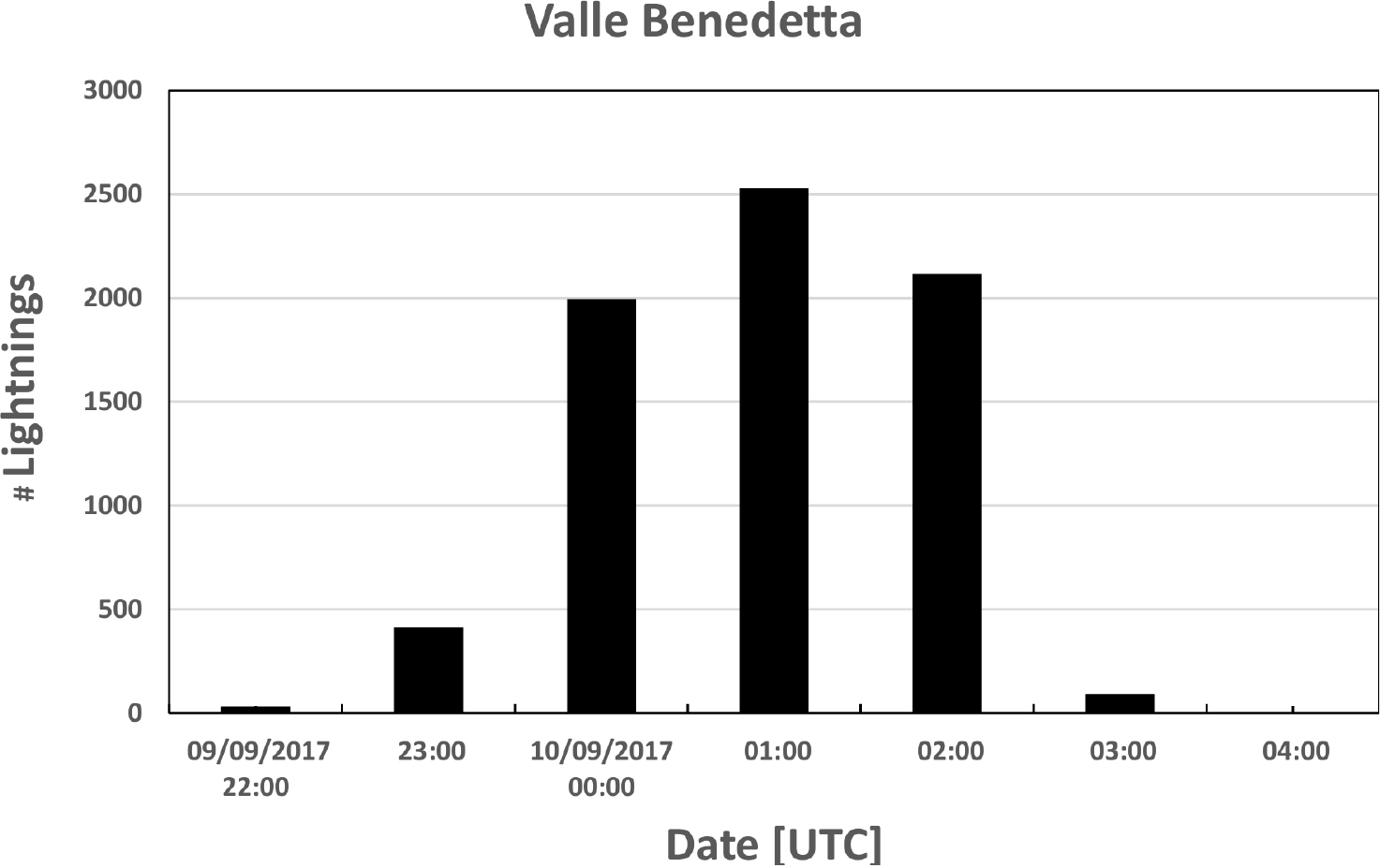}\label{fig:5a}}~~
 \subfloat[]{\includegraphics[width=0.46\textwidth]{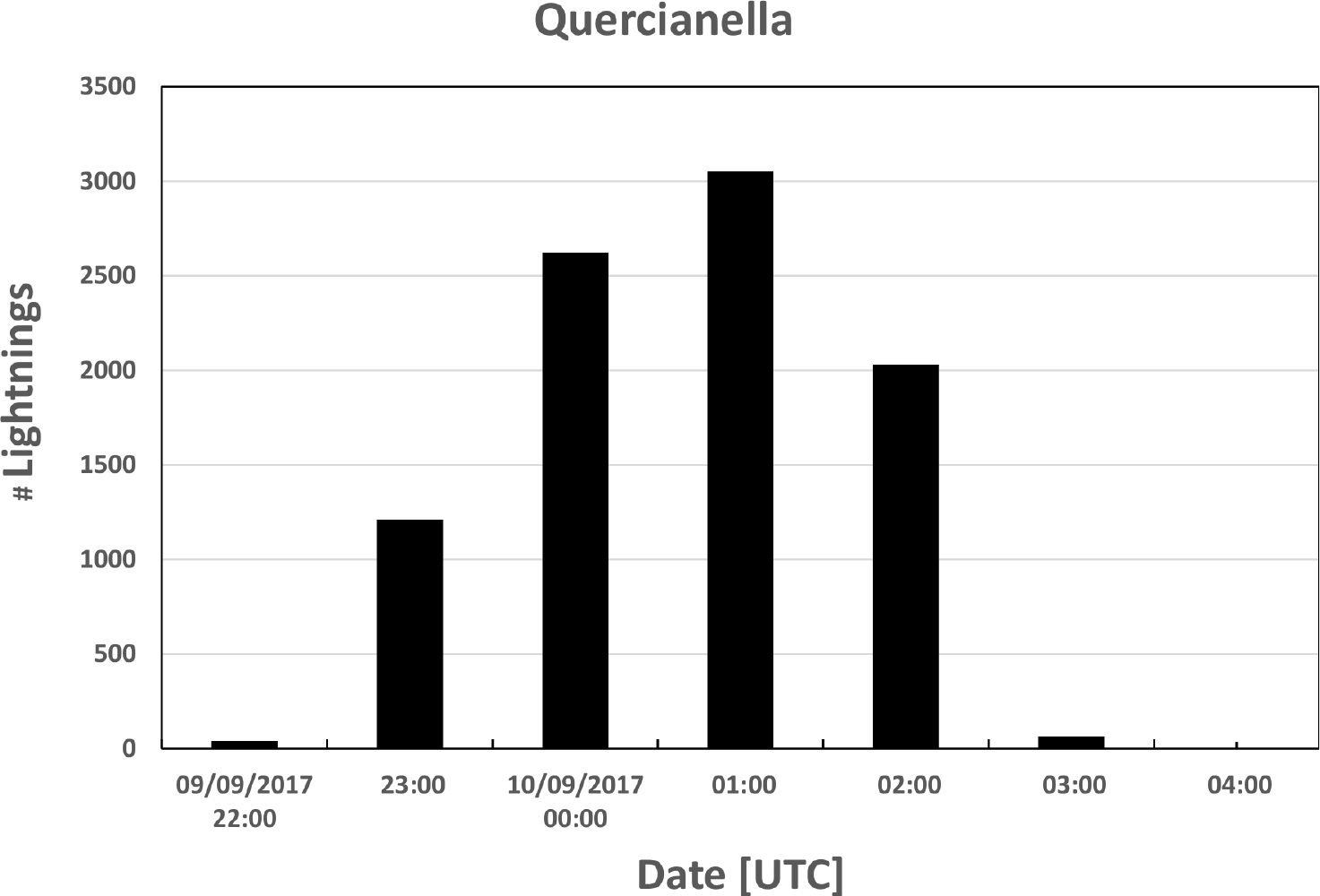}\label{fig:5b}}\\
 \caption{(\textbf{a}): Map of lightning strikes measured between 2150 UTC, 9th of September 2017, and 0350 UTC, 10th of September 2017. (\textbf{b}): histogram of the number of strokes per hour within $\pm0.1$ degree of the Valle Benedetta rain-gauge, registered from 2200 UTC of the 9th of September to 0400 UTC of the 10th of September 2017. (\textbf{c}): as in (\textbf{b}) but for the Quercianella rain-gauge. Data were taken from the Blitzortung dataset (www.blitzortung.org/, link accessed in April 2021).}
\label{figure_three}
\end{figure}

The deep convection of the Livorno event produced (see Figure \ref{figure_three}a) an average of four lightning strikes per second over Northern Tuscany in the six hours  between 2150 UTC of the 9th of September and 0350 UTC of the 10th of September; this number is approximately 10\% of the global flash rate \cite{christian2003global}.
For those  rain-gauges that recorded the greatest amount of precipitation (Valle Benedetta and Quercianella), corresponding lightning time series graphs are shown in Figures \ref{figure_three}b and \ref{figure_three}c as histograms of number of strokes per hour within $\pm$0.1 degree of the two weather stations centres, from 2200 UTC of the 9th September to 0400 UTC of the 10th of September 2017. A total number of 9015 and 7177 strokes were detected during the 6-hour period for Quercianella and Valle Benedetta stations, respectively. The highest number of flashes was recorded at 0100 UTC of the 10th of September for both Quercianella and Valle Benedetta (3053 and 2528, respectively).
%
\subsection{Modelling setup}
The numerical model used in this work to simulate the 9-10 September Livorno case is the WRF model \citep{skamarock2019description}. The Mesoscale and Microscale Meteorology (MMM) Laboratory at the National Center for Atmospheric Research (NCAR) has led the development of the WRF model since its inception in the late 1990s. It is a fully compressible, Eulerian, non-hydrostatic mesoscale model, designed to provide accurate numerical weather forecasts both for research activities and operations. In this work, we implemented the Advanced Research WRF (ARW) version of the model updated to version 4.0 (June 2018). The model dynamics, equations and numerical schemes implemented in the WRF-ARW core are fully described in \cite{skamarock2019description} and \cite{klemp2007conservative}, while the model physics, including the different options available, is described in \cite{chen2000annual}. A summary of the model settings used in this study is given in Table \ref{tab:wrf}, while the domain of integration is indicated in Figure \ref{fig:syn} with the red rectangle. All the WRF simulations (with and without assimilation) performed in this study share the same basic set of physical parameterisations listed in Table \ref{tab:wrf}. This configuration mirrors the one that was operational at the weather service of the Tuscany regional (LaMMA, see www.lamma.toscana.it, link accessed in April 2021) at the time of the Livorno event, although in this study we used a more updated WRF model version (4.0 instead of 3.9).
\begin{table}[H]
\caption{Basic settings of the WRF model simulations.}\label{tab:wrf}
\centering
\begin{tabular}{l l}
\hline
\textbf{Variable} & \textbf{Value} \\
\hline
Rows$\times$Columns & 440$\times$400 \\
Vertical levels & 50 \\
Grid spacing &  3 km \\
Time step &  12 s \\
Cumulus convection &  explicit (no parameterisation) \\
Micro-physics option &  Thompson scheme \cite{thompson2014study} \\
Boundary-layer option &  Yonsei University \citep{hong2006new} \\
Land-surface option &  Unified Noah model \citep{chen1996modeling} \\
\hline
\end{tabular}
\end{table}
To start the WRF simulations, initial and boundary conditions are derived from the ECMWF-IFS global model data operational at the time of the event (model cycle 43r3). The spectral resolution is TCO1279, (where CO means that a cubic-octahedral grid is used), which corresponds to a grid spacing of approximately 9 km; the number of vertical levels is 137 and the top of atmosphere is set to 0.01 hPa. 

The control forecast is initialized at 1200 UTC on the 9th of September 2017 and the forecast length is set to 15 hours (ending time is 0300 UTC of the 10th of September). To test the impact of assimilating conventional and radar data, we utilised the WRF data assimilation (WRFDA) system \citep{barker2012weather}. The WRFDA software was used to perform a 3-hour cycling 3D-Var data assimilation using the rapid update cycle approach, similar to what is found in \cite{Schwitalla2014review}. We implemented an assimilation step every 3 hours starting from 1200 UTC on the 9th of September up to 1800 UTC of the same day with an assimilation window of $\pm$ 30 minutes, i.e., all the  observations between $t-30$ minutes and $t+30$ minutes are assumed to be valid at the analysis time $t$. The 3D-Var implementation of the WRFDA package relies on previous developments designed for the fifth-generation Penn State/NCAR Mesoscale Model \cite{barker2004three}. Although detailed descriptions of the 3D-Var method can be found in \cite{xiao2005assimilation,xiao2007multiple,wang2013indirect}, the technique can be summarised as follows: the basic idea is to estimate the optimal state of the atmosphere (in the model space) by using the observations available, a short-term forecast (often referred as first guess or background) and information about errors statistics on both the observations and the model. Let $t\in \mathbb{R}$ be the time of the analysis and let $\mathbf{x} = \mathbf{x}(t) \in \mathbb{R}^n$ the model analysis at time $t$. Using an iterative process, the 3D-Var method looks for the minimum value of the cost function $J(\mathbf{x})$, defined as:
\begin{equation}\label{eq:3dvar}
J(\mathbf{x}) = \frac{1}{2}\left\{ 
                 (\mathbf{x} - \mathbf{x_b})^{T} \mathbf{B}^{-1}  (\mathbf{x} - \mathbf{x_b}) +
                 (\mathbf{y_o} - H(\mathbf{x}))^{T} \mathbf{R}^{-1} (\mathbf{y_o} - H(\mathbf{x})) 
                 \right\},
\end{equation}
where, following the notations of \cite{courtier1998ecmwf}, $\mathbf{x_b}\in \mathbb{R}^n$ is the background model state at time $t$, $\mathbf{B}\in \mathbb{R}^{n\times n}$ is the covariance matrix of the background errors, $\mathbf{y_o}\in \mathbb{R}^p$ is the vector of observations ($p < n$) at time $t$, $H:\mathbb{R}^n\mapsto\mathbb{R}^p$ is the  observation operator that transforms the variables from the model state to the observation space and $\mathbf{R}\in \mathbb{R}^{p\times p}$ is the covariance matrix of observation errors.

Accurate estimates of the $\mathbf{B}$ and $\mathbf{R}$ matrices determine the quality of the analysis. As regards the $\mathbf{R}$ matrix, default values provided by the WRFDA software (see User's Guide \cite{skamarock2019description}) were used for diagonal elements, whereas off-diagonal's elements are set to zeros. In fact, correlations between different instruments are usually assumed as null. Although some studies demonstrated that including inter-correlations in the $\mathbf{R}$ matrix may provide better analyses \cite{liu2003potential}, this mainly holds when satellite radiance data are considered \cite{bormann2010estimates,bormann2011estimates}, thus we claim that the assumption on the $\mathbf{R}$'s structure is valid in our study. We computed the background error correlation matrix $\mathbf{B}$ by means of the National Meteorological Center
(NMC) method \cite{parrish1992national}, which estimates the value of the elements of the matrix statistically, by averaging the differences between two short-term forecasts valid at the same time but initialised one shortly after the other (we set 12 hours later). The $\mathbf{B}$ matrix we used is the result of the NMC method applied for ten months (from December 2018 to October 2019).

Following \cite{sun1997dynamical}, the observation operator $H$ for the radial velocity $V_r$ is defined by the relationship:
\begin{equation}\label{eq:radvel}
V_r=u\frac{x-x_i}{r_i}+v\frac{y-y_i}{r_i}+(w-v_T)\frac{z-z_i}{r_i}\ \ \ ,
\end{equation}
where $(u,v,w)$ are the modelled wind components, $(x_i,y_i,z_i)$ are the coordinates of the radar antenna, $(x,y,z)$ are the coordinates of the radar observation, $r_i$ is the distance between $(x,y,z)$ and $(x_i,y_i,z_i)$ and $v_T$ is the mass-weighted terminal velocity of the precipitation. To calculate $v_T$, the default formula \cite{sun1998dynamical} was used, that is:
\begin{equation}\label{eq:termvel}
v_T=5.4\cdot(p_0/\Bar{p})^{0.4}\cdot q_r^{0.125}\ \ \ ,
\end{equation}
where $p_0$ is the surface pressure, $\Bar{p}$ is the base-state pressure and $q_r$ (unit g$\cdot$ kg$^{-1}$) is the model predicted rainwater mixing ratio. Let $Z$ be the reflectivity data expressed in dBZ, then the nonlinear $Z-q_r$ relationship is defined by \cite{sun1997dynamical}:
\begin{equation}\label{eq:h}
Z=c_1+c_2\cdot \log_{10}(\rho q_r)\ \ \ ,
\end{equation}
where $c_1=43.1$ and $c_2=17.5$ are constants and $\rho$ (unit kg$\cdot$ m$^{-3}$) is the air density. Following the performances achieved by \cite{lagasio2019predictive}, radar reflectivity data were assimilated using the indirect technique proposed by \cite{wang2013indirect}, that is by inverting the $Z-q_r$ relation given in Equation \ref{eq:h} and assimilating observed rainwater mixing ratio estimated from reflectivity values. 

Before assimilation, radar reflectivity data undergo to a thinning procedure. This consists in eliminating ground/sea clutter and georeferencing radar data onto the model grid; then for each model grid point the procedure selects the maximum reflectivity value (and the corresponding radial velocity) among radar observations belonging to that grid point. Finally, to reduce the spin-up in the first hours of integration, we implemented the digital diabatic filter initialization \cite{peckham2016implementation} at each starting time of the WRF simulationsw.

\begin{figure}[H]
\centering
\includegraphics[width=\textwidth]{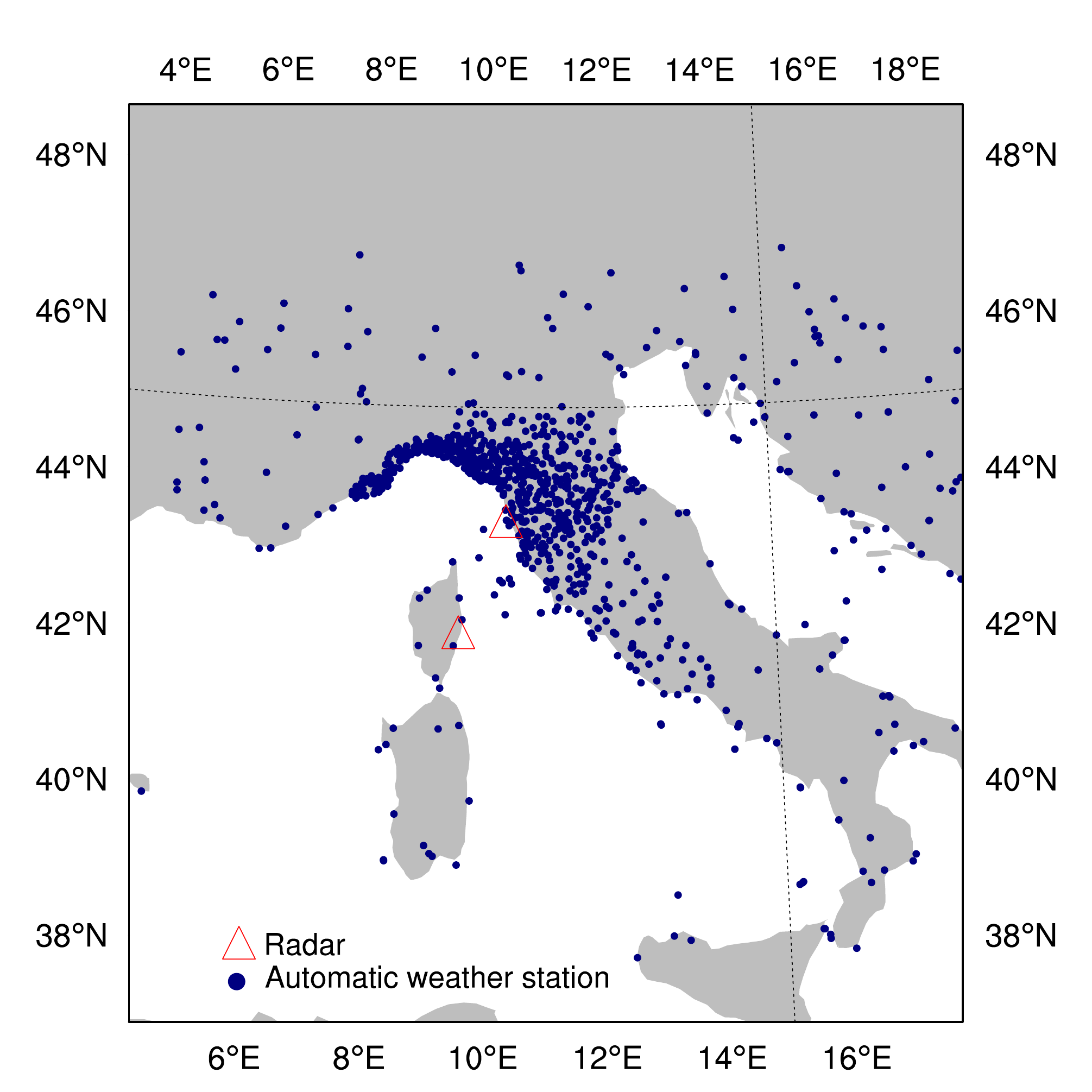}
\caption{Geographical extent of the domain of integration of the WRF model. The two red triangles ``\textcolor{red}{$\bigtriangleup$}'' indicate the locations of the Aleria (in the Corsica region) and Livorno (on the Tyrrhenian coast) radars. The blue dots indicate the locations of the ground weather stations assimilated during the assimilation steps.}
\label{fig:WRFdomain}
\end{figure}

To assess the impact of assimilating different types of observations, three experimental forecasts have been deployed and their results are compared with those obtained with the control forecast (named with the code $C$), which doesn't assimilate any observational dataset. The first experimental forecast (code $S$) assimilates data from ground weather stations, the second (code $S+R$) assimilates radar data in addition to ground weather station data. Conventional data assimilated in both experiments are: pressure, temperature and relative humidity at 2-metre above ground level, wind speed and direction at 10-metre above ground level. The total number of weather stations reporting valid data is 1036 for the assimilation step performed at 1500 UTC on the 9th of September and 1051 for the 1800 UTC assimilation. Remote sensed data assimilated are reflectivity data from the X- and S-band radars and radial velocity data from the Aleria S-band radar. The locations of weather stations and radars are indicated with the blue points and red triangles, respectively, in Figure \ref{fig:WRFdomain}.

To investigate the role of sea surface temperatures (SSTs) in the Livorno case, we performed a further numerical experiment (code $S+R+M$), which shares the same settings and design of the $S+R$ experiment, but includes a simplified ocean model that modifies its status (SSTs and ocean mixed layer depth) thanks to the interface heat, momentum and mass fluxes. In fact, the interaction between atmosphere and ocean can play a key role in the formation and intensification of extreme atmospheric events \cite{foussard2019response} through the energy fluxes, which can impact the generation of HPEs by modifying the structure of the boundary layer, the distribution of wind fields and therefore the position of the convergence line \cite{chelton2004satellite,o2003observations}. Historically, numerical weather models implemented the SSTs as a static boundary, taken from the global models or from  low-resolution satellite datasets. However, following this approach it is difficult to reproduce complex energy feedback between air and sea \cite{ricchi2016use}, in particular in the coastal areas. In order to limit this impact, without slowing down the calculation time with the coupling of an ocean model, we proceeded by using a ``slab ocean'', also known as a simple Ocean Mixed Layer \cite{davis2008prediction}, which updates the SSTs every hour, according to the energy fluxes at the air-sea interface. This approach is based on a simplified model that aims at implementing the SSTs and ocean mixed layer depth into the WRF model. This one-dimensional approach is initialized with SST data, a mixed layer depth value of 40 m and a lapse rate 0.14  K $\cdot$ m$^{-1}$. These data are provided by the Copernicus Marine Services (CMEMS) numerical model. WRF modifies the SSTs value according to the energy and momentum fluxes at the interface, not taking into account the ocean dynamics, but making each grid point of the evolve as a function of the energy budget at surface.

\begin{figure}[H]
\centering
\includegraphics[width=0.99\textwidth]{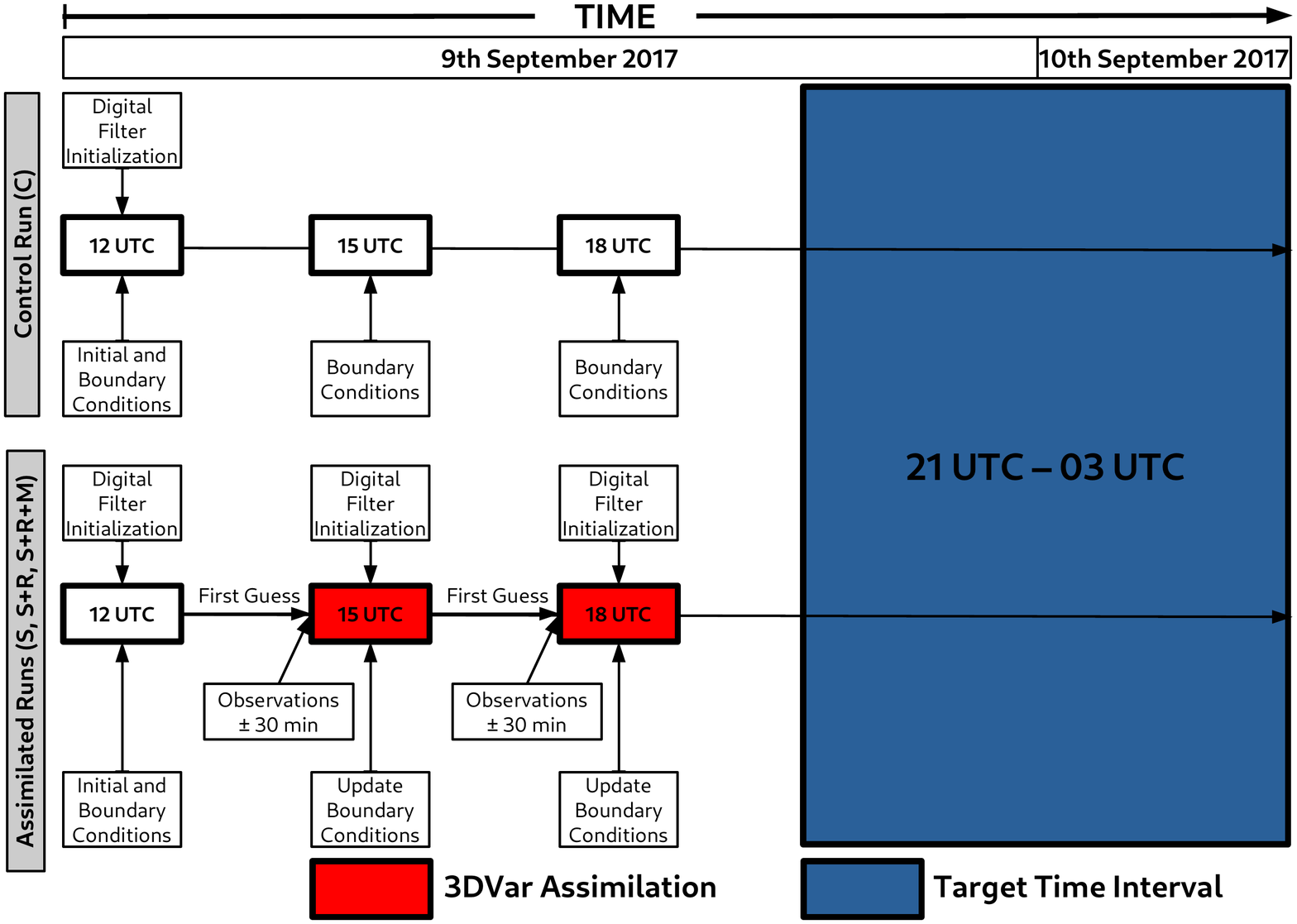}
\caption{Modelling setup of the WRF control run $C$ and assimilated runs $S$, $S+R$ and $S+R+M$.}
\label{figure_scheme}
\end{figure}

Table \ref{tab:set-ups} shows a summary of the various forecasts and the codes adopted to name them, while Figure \ref{figure_scheme} shows a scheme of the modelling setup.
\begin{table}[H]
\caption{Codes assigned to name the forecasts and description of the data used in each assimilated forecast.}
\label{tab:set-ups}
\centering
\begin{tabular}{ll}
\toprule
\textbf{Forecast code}	& \textbf{Data assimilated}\\
\midrule
$C$     & none (control run) \\
\midrule
$S$		& Convectional data from weather stations:\\
		& pressure, 2-metre temperature\\
		& 2-metre relative humidity\\
		& 10-metre wind speed and direction\\
\midrule
$S+R$	& as in $S$ plus reflectivity data from X- and S-band radars,\\
		& and radial velocity data from the S-band radar\\
\midrule
$S+R+M$	& as in $S+R$ \\
        &      (it differs from the $S+R$ experiment because \\
        &       it implements a simplified marine model)\\
\bottomrule
\end{tabular}
\end{table}
%
\section{Quantitative precipitation forecast verification}\label{sec:pacman}
When evaluating the performance of different forecasts, a quantitative evaluation of the spatial agreement between predicted and observed values is crucial. A direct numerical comparison can be misleading, especially for variables whose spatial distribution is highly concentrated in a small area, as happens during HPEs. In fact, any forecast that correctly predicts the occurrence of a highly localized heavy rain, may incur in the so called ``double penalty'' error \citep{ebert2009neighborhood} if it places the event in a nearby area, producing, for example, a root mean squared error ($RMSE$, see \cite{wilks2011statistical}) higher than another forecast which completely misses the prediction. To overcome such a limitation, object-based verification methods have been developed by the scientific community \cite{Davis2006_Obj-BasedVerification1} and, currently, software packages for practical applications exist \cite{ModelEvaluationToolsSite}. Anyway these methods exhibit some drawbacks, like the smoothing and filtering the observations undergo and the large number of parameters whose setting turns out somewhat arbitrary. 

Considering the features of the phenomenon under examination, we have devised a simple and robust \textit{ad-hoc} verification method, which does not alter in any way the recorded rainfall values. No arbitrary or subjective parameters are required, except for the ones needed to define the area where the event occurred, namely the center and radius of the circle in which the area of interest is assumed inscribed. For the flash flood under examination, we chose a circle with radius 0.4 degrees in the longitude-latitude forecast domain and centered at the point $P_C$ with coordinates $(x_{C,lon},x_{C,lat})=( 10.64^{\circ}\text{E},43.75^{\circ}\text{N})$, about 30 km to the North-East of the  Livorno town. This area includes 91 rain-gauges located at points $P_i$ ($i=1,\ldots,91$), which measured the highest rain rate in the 6-hour period ending on 0300 UTC of the 10th of September (see Figure \ref{figure_one}). By applying affine transformations\footnote{More precisely rigid  roto-translations preserving the Euclidean distance; these transformations are usually denoted as the $SE(2)$ group.} to the 91 position vectors $\mathbf{x}_i=(P_i-P_C) \in \mathbb{R}^2$, we searched
\begin{figure}[H]
\centering
\includegraphics[height=\textwidth,angle=-90]{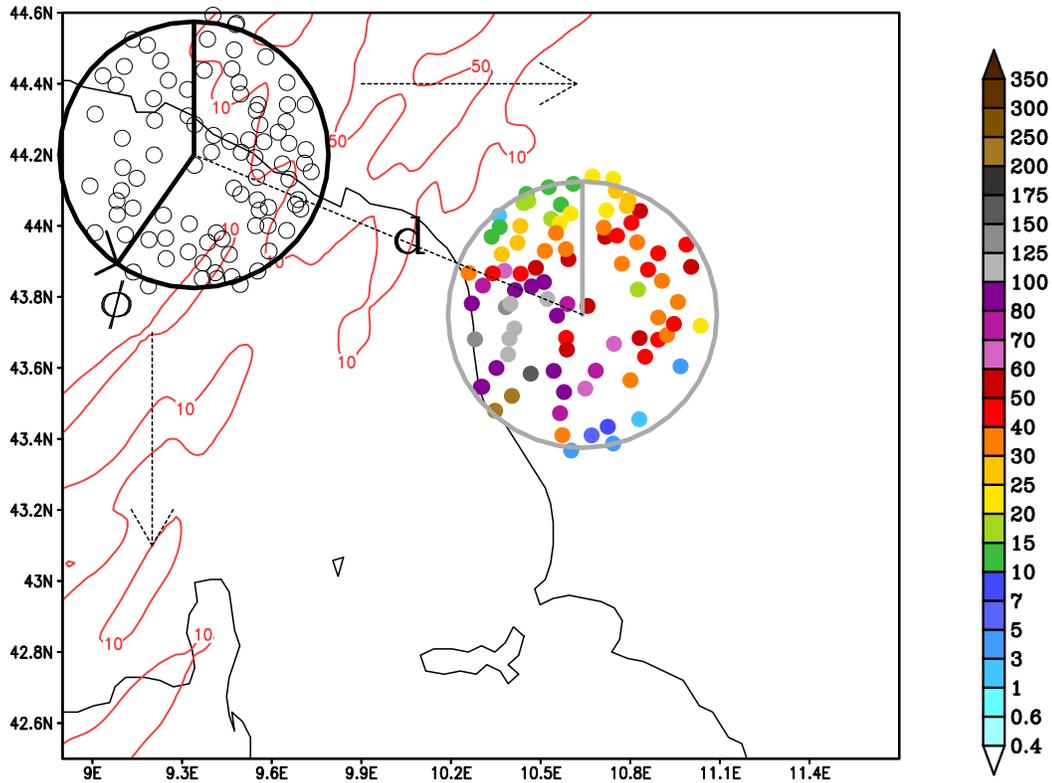}
\caption{The gray circle contains the 91 rain-gauges (coloured points) that registered the heaviest rainfall in the 6-hour ending at 0300 UTC on the 10th of September. The sample black circle contains the transformed (i.e., roto-translated) 91 rain-gauges obtained by varying the parameters $d$ and $\phi$. For each forecast, the verification procedure extracts the predicted precipitation values (samples shown  with the red contours) at the black point locations and  calculates the root mean squared error against values reported at the coloured points.}\label{fig:pacman_method}
\end{figure}
for the transformation minimising the $RMSE$ between the cumulative rain $r_i$ measured at station $P_i$ and the one $f(\mathbf{y}_i)$ predicted by the forecast model $f$ at the transformed points $\mathbf{y}_i\in \mathbb{R}^2$ (see Figure \ref{fig:pacman_method}). Since we applied a roto-translation, any point $\mathbf{y}_i$ is determined by:
\begin{equation}\label{eq:rototra}
    \mathbf{y}_i=R_{\phi}(\mathbf{x}_i)+\mathbf{x}_0\ \ \ ,
\end{equation}
where $\phi$ is the rotation angle, $R_{\phi}\in \mathbb{R}^{2\times2}$ is the rotation operator and has the form:
\begin{equation}
R_{\phi}=\begin{bmatrix} \cos \phi & - \sin\phi \\ \sin\phi & \cos \phi \end{bmatrix},    
\end{equation}
and $\mathbf{x}_0=(x_{0,lon},x_{0,lat})\in \mathbb{R}^2$ is the translation vector. All the parameters of the roto-translation are defined in the longitude-latitude space. We note that the distance, in km, between $P_C$ and the centre of any circle translated by a vector $\mathbf{x}_0$ is given by:
\begin{equation}\label{eq:d0}
d_0\simeq R\sqrt{\cos^2(x_{C,lat})\cdot x_{0,lon}^2+x_{0,lat}^2}\ \ \ ,
\end{equation}
where $x_{0,lon}$, $x_{0,lat}$ and $x_{C,lat}$ are expressed in radians and $R$ is the Earth radius (approximately 6370 km). The minimisation of the positive defined error function
\begin{equation}\label{eq:rmse}
{RMSE}=\sqrt{\langle (f(\mathbf{y}_i)-r_i)^2 \rangle}\ \ \ ,
\end{equation}
where $\langle \cdot \rangle$ stands for the average operator over the index $i$, determines the particular set of parameters for which the agreement between predicted and observed cumulative rain is the highest possible, at least in terms of $RMSE$. In this way, the location and intensity errors are disentangled, the former given by the affine transformation parameters amplitude ($\phi$ and $\mathbf{x}_0$), the latter by the minimised $RMSE$.

To contend with the validation method described above, standard verification scores were computed. Observed rainfall data registered at the 91 rain-gauges were compared with corresponding modelled data, which were extracted at the 91  nearest grid points containing rain-gauge locations. Standard verification statistics considered are: $RMSE$ (defined in Equation \ref{eq:rmse}), mean error ($ME$) and the multiplicative bias  ($mbias$). Let  $r_i$ and $f(P_i)$ be the observed and modelled rainfall values at location $P_i$, respectively,  then $ME$ and $mbias$ are defined as follows (see also \cite{wilks2011statistical}):
\begin{align}
ME    = & {\langle f({P}_i)-r_i\rangle},\label{eq:me}\\ 
mbias = &\frac{\langle f({P}_i)\rangle}{\langle r_i\rangle}.\label{eq:mbias}
\end{align}

Furthermore, to summarise multiple aspects of the model performance in a single diagram, we made use of the performance diagrams (see \cite{roebber2009visualizing}). Such diagrams plot four skill measures of dichotomous (yes/no) forecasts: probability of detection ($POD$), success ratio ($SR$), frequency bias ($bias$) and critical success index ($CSI$). Using the $2\times2$ contingency table shown in Table \ref{tab:contingency}
\begin{table}[H]
\caption{The $2\times2$ contingency table.}\label{tab:contingency}
\begin{center}
\begin{tabular}{l|lcc}
\hline
\multirow{2}{*}{}                                   &     & \multicolumn{2}{c}{Event Observed} \\\hline
                                                    &     & yes              & no              \\
\multicolumn{1}{c|}{\multirow{2}{*}{Event Forecast}} & yes & A                & B               \\
\multicolumn{1}{c|}{}                                & no  & C                & D              \\\hline
\end{tabular}
\end{center}
\end{table}
the four skill measures are defined as follows:
\begin{eqnarray*} 
POD  = & \frac{A}{A+C},\\ 
SR   = & 1 - \frac{B}{A+B},\\
bias = & \frac{A+B}{A+C},\\
CSI  = & \frac{A}{A+B+C}.
\end{eqnarray*}
%
\section{Results}\label{sec:results}
\begin{figure}[H]
\centering
\includegraphics[height=\textwidth,angle=-90]{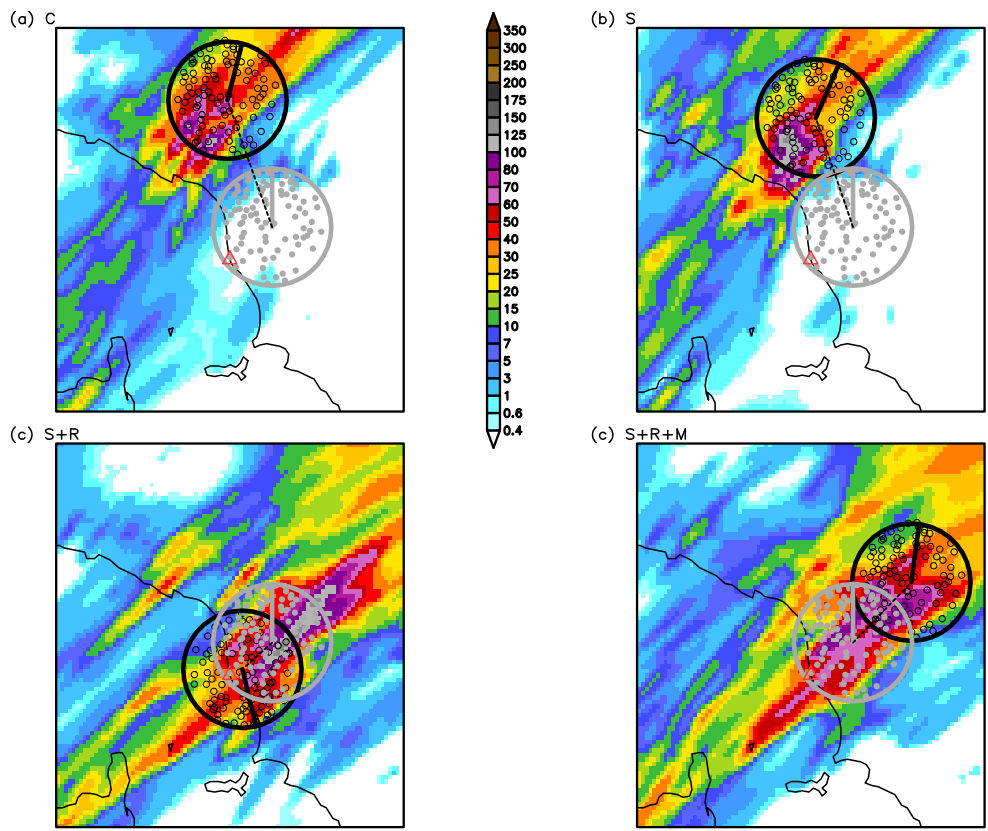}
\caption{Rainfall forecast accumulated in the 6-hour period ending on 0300 UTC of the 10th of September 2017 (shaded colours, unit mm): (\textbf{a}): control run $C$, (\textbf{b}): assimilated run $S$, (\textbf{c}): assimilated run $S+R$, (\textbf{d}): assimilated run $S+R+M$. The black circles are the outputs of the roto-translations of the gray circle, minimising the $RMSE$ of the rainfall with respect to the values measured by the 91 rain-gauges covering the area mostly affected by the heaviest precipitations (rain-gauge positions indicated in gray). The dashed segment indicates the distance between the centres of the black and gray circles and its length is reported in Table \ref{tab:rottra} ($d_0$ values). The solid black segment shows the rotation of the circle with respect to the North direction, counterclockwise. The red triangle (``\textcolor{red}{$\bigtriangleup$}'') indicates the Livorno town.}\label{pacman}
\end{figure}

In Figure \ref{pacman}, we show with the shaded colours the quantitative precipitation forecast (QPF) in the 6-hour period ending at 0300 UTC of the 10th of September for the four experiments listed in Table \ref{tab:set-ups}.
In each panel, the gray circle indicates the area containing the 91 rain-gauges (closed gray points) that registered the highest precipitations. The visual inspection of Figure \ref{pacman} indicates that the $C$ and $S$ forecasts provide QPF values close to zero in the Livorno area. On the other hand, both numerical experiments predict QPF peaks up to 100-120 mm in an area few tens of kilometres (approximately 40-60 km) to the North of the area of interest, where rain-gauges registered accumulated precipitations up to 30-40 mm (see Figure \ref{fig:1a}). Both the $S+R$ and $S+R+M$ forecasts predict high QPF values inland of the Livorno area, however, QPF maxima (in the range 125-150 and 80-100 mm for the $S+R$ and $S+R+M$ forecast, respectively) largely under-estimate actual observed values, which reached approximately 200-230 mm (see Figure \ref{fig:1b}). 

In Table \ref{tab:verif}, we show the verification skill scores defined in Equations \ref{eq:rmse}, \ref{eq:me} and \ref{eq:mbias}. The scores were computed using observed rainfall registered at the 91 rain-gauges shown in the gray circle of Figure \ref{pacman} and the corresponding modelled precipitations extracted at the grid points closest to the rain-gauge locations.
\begin{table}[H]
\caption{Verification scores defined in Equations \ref{eq:rmse}, \ref{eq:me} and \ref{eq:mbias} for the four experimental runs.}\label{tab:verif}
\centering
\begin{tabular}{lllllll}
\toprule
\textbf{Forecast code}	& {$RMSE$} & {$ME$} & $mbias$\\
\midrule
$C$     & 65.8 & -50.3 & $<10^{-3}$ \\
$S$     & 65.9 & -50.5 & $<10^{-3}$ \\
$S+R$   & 46.3 & -2.5  & 0.95 \\
$S+R+M$ & 40.4 & -2.9  & 0.94 \\
\bottomrule
\end{tabular}
\end{table}
The $S+R$ and $S+R+M$ forecasts exhibit the more satisfactory scores; in fact the $RMSE$s are approximately 40/45 mm, whereas $C$ and $S$ errors are higher and reach approximately 65 mm. The $ME$s and biases demonstrate the underestimation of all the four  predictions; such underestimation  is remarkable for $C$ and $S$ ($ME$ $\simeq$ -50 mm and $mbias<10^{-3}$). On the other hand, the $S+R$ and $S+R+M$ forecasts provide more skillful scores, with the $mbias$ close to the perfect score 1 and $ME$s approximately -2.5/-3 mm.

\begin{figure}[H]
\centering
\includegraphics[height=0.99\textwidth,angle=0]{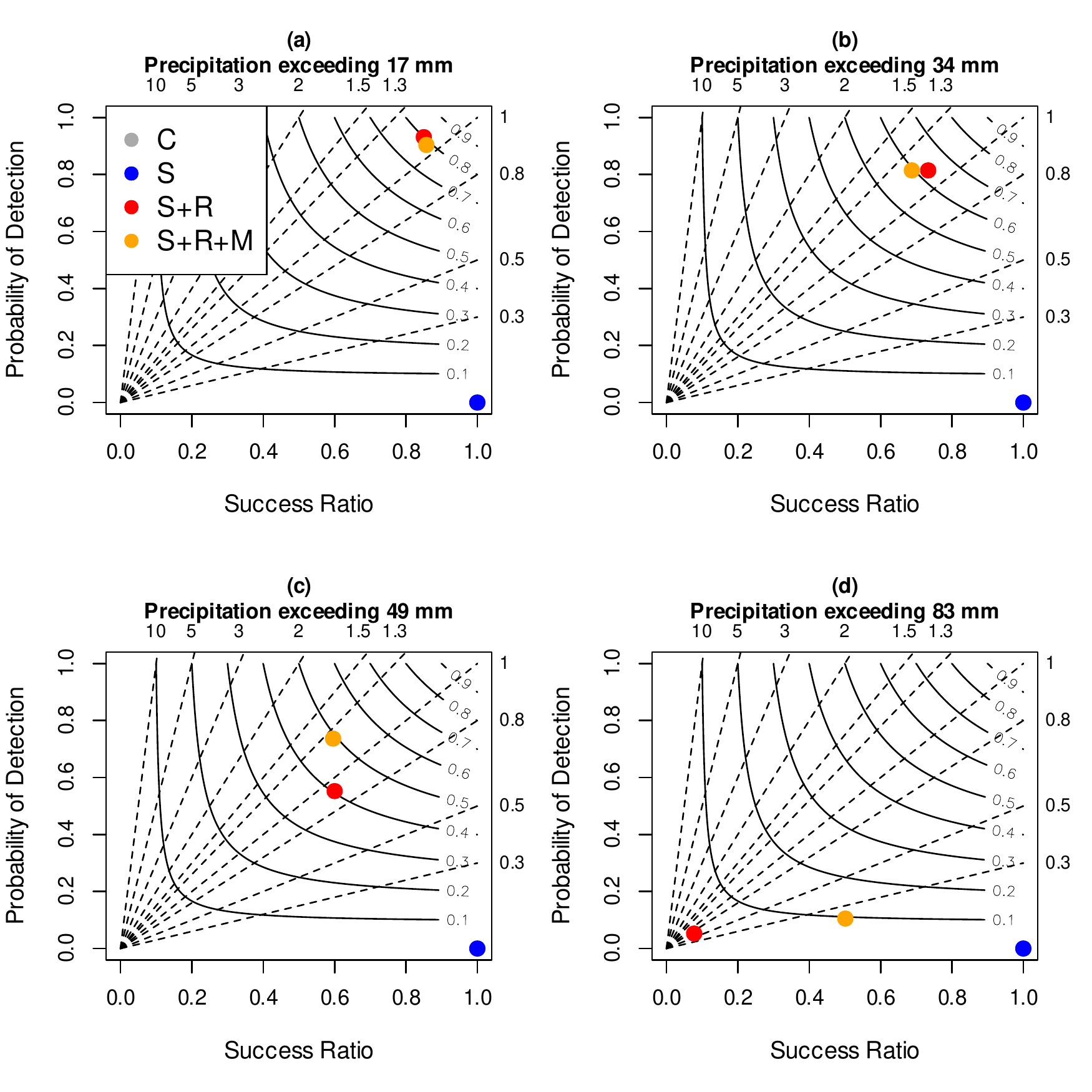}
\caption{Performance diagrams for selected precipitation thresholds accumulated in the 6-hour period ending at 0300 UTC of the 10th of September. In each panel the X-axis shows the Success Ratio ($SR$), the Y-axis shows the Probability of Detection ($POD$), the curved lines represent Critical Success Index ($CSI$) values, and the dashed diagonal lines represent $bias$. In each panel, the blue point overlaps the gray point.}
\label{fig:perfdiag}
\end{figure}

In Figure \ref{fig:perfdiag}, we show the performance diagram for selected rainfall thresholds corresponding to approximately the 20th, 40th, 60th and 80th percentiles of the observational dataset. The plots show that, for precipitation thresholds equal to 17, 34 and 49 mm, the $S+R$ and $S+R+M$ forecasts behave similarly and outperform the $C$ and $S$ forecasts, which have no skill since both matching points lie close to the bottom-right corner. As the precipitation threshold equals 83 mm, which approximately corresponds to the 80th percentile of the observations, the $S+R$ forecast doesn't provide any valuable information, since the matching point lies close to the bottom-left corner. The $S+R+M$ forecast holds some skill as regards the $bias$ (approximately 0.5) and the $SR$ score (approximately 0.3), whereas the skill of $CSI$ ($\simeq10$) and $POD$ ($<0.2$) is limited. Skill scores for precipitations thresholds greater than the 80th percentile of the observations are poorer or null  for all the forecasts (performance diagrams not shown).

In each panel of Figure \ref{pacman}, we also show the result of the roto-translation of the gray circle minimising the $RMSE$ between the observed and predicted rainfall data (the transformed circle and rain-gauge locations are indicated with the black colour). The angle of rotation $\phi$ is measured, by definition, counterclockwise from the North direction. For each experimental run, the parameters of the $RMSE$ minimising  roto-translation are shown in Table \ref{tab:rottra}.
\begin{table}[H]
\caption{Parameters of the roto-translation minimising the $RMSE$ between observed and predicted values extracted at the 91 rain-gauge locations located within the gray and black circles, respectively, shown in Figure \ref{pacman}. {$RMSE_{min}$} is the error after the minimising roto-translation; $x_{0,lon}$ and $x_{0,lat}$ are the coordinates of the translation vector $\mathbf{x}_0$ defined in Equation \ref{eq:rototra}; $d_0$ indicates the distance, in km, between the centres of the black and gray circles (i.e, the dashed segment of Figure \ref{pacman}) and is given in Equation \ref{eq:d0}. The angle $\phi$ indicates the rotation amplitude with respect to the North direction, counterclockwise.}\label{tab:rottra}
\centering
\begin{tabular}{lllllll}
\toprule
\textbf{Forecast code} & {$RMSE_{min}$} & $x_{0,lon}$ & $x_{0,lat}$ &$d_0$ & $\phi$\\
\midrule
$C$     & 29.7  & -0.36 & +0.87 & 100 & $343^\circ$\\
$S$     & 29.0  & -0.30 & +0.75 & 86  & $333^\circ$\\
$S+R$   & 29.9  & -0.24 & -0.18 & 27  & $196^\circ$\\
$S+R+M$ & 30.1  & +0.48 & +0.41 & 60  & $351^\circ$\\
\bottomrule
\end{tabular}
\end{table}
After the roto-translations, all the four experiments exhibit similar {$RMSE_{min}$}, with the experiment $S$ providing the lowest value (29.0 mm) and experiment $S+R+M$ having the largest one (30.1 mm). However, the parameter $d_0$ indicates that the $S+R$ experiment displaces the transformed black circle by approximately 27 km, whereas the other experiments provide higher displacements, ranging from 60 to 100 km. We stress the fact that not only the displacement is lower when radar data are assimilated, but also oriented along the axis of the perturbation hitting the Livorno town (from South-West to North-East). The angle of the rotation $\phi$ reported in Table \ref{tab:rottra} spans from $196^\circ$ for $S+R$ to $351^\circ$ for $S+R+M$.

\begin{figure}[H]
\centering
\includegraphics[height=0.99\textwidth,angle=-90]{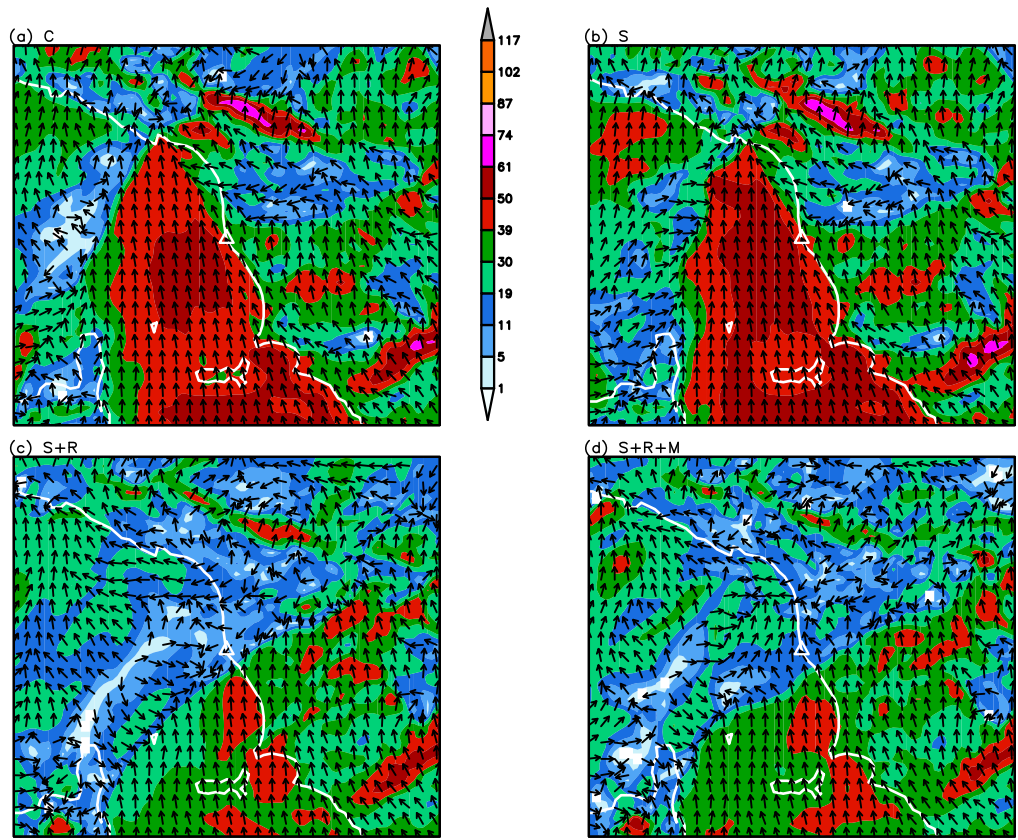}
\caption{10-metre wind speed  (unit km$\cdot$ h$^{-1}$) and direction at 0000 UTC of the 10th of September for (\textbf{a}): control run $C$, (\textbf{b}): assimilated run $S$, (\textbf{c}): assimilated run $S+R$, (\textbf{d}): assimilated run $S+R+M$. The white triangle indicates the Livorno town.}
\label{figure_wind}
\end{figure}

To further evaluate the impact of assimilating conventional and radar data, in Figure \ref{figure_wind} we show the 10-metre wind speed and direction at  0000 UTC on the 10th of September for the four numerical experiments. All the four forecasts reconstruct a well defined convergence line over the Ligurian Sea between southerly and westerly winds, which is responsible for triggering convective rainfall. However, in the $C$ experiment, such convergence line is positioned few tenths of kilometres (approximately 70 km) to the North of the Livorno area, causing the wrong displacement of precipitations. The impact of the conventional data (experiment $S$) is negligible and doesn't modify the position of the convergence line with respect to the control run $C$ (see panels (a) and (b) in Figure \ref{figure_wind}). On the other hand, assimilating radar data modifies substantially the 10-metre wind speed and direction forecast, with a better localization of the convergence line, positioned close to the area that registered the heaviest precipitations (see panel (c) in Figure \ref{figure_wind}). The implementation of the simplified ocean model (see panel (d) in Figure \ref{figure_wind}) generates prefrontal winds (southern part of the convergence line) more intense of about 8 m/s, and a local pressure decrease of about 1.5 hPa (map not shown); as a consequence, the front-genesis evolves more quickly with greater intensity in the event area.

\begin{figure}[H]%
 \centering
 \includegraphics[height=0.99\textwidth,angle=-90]{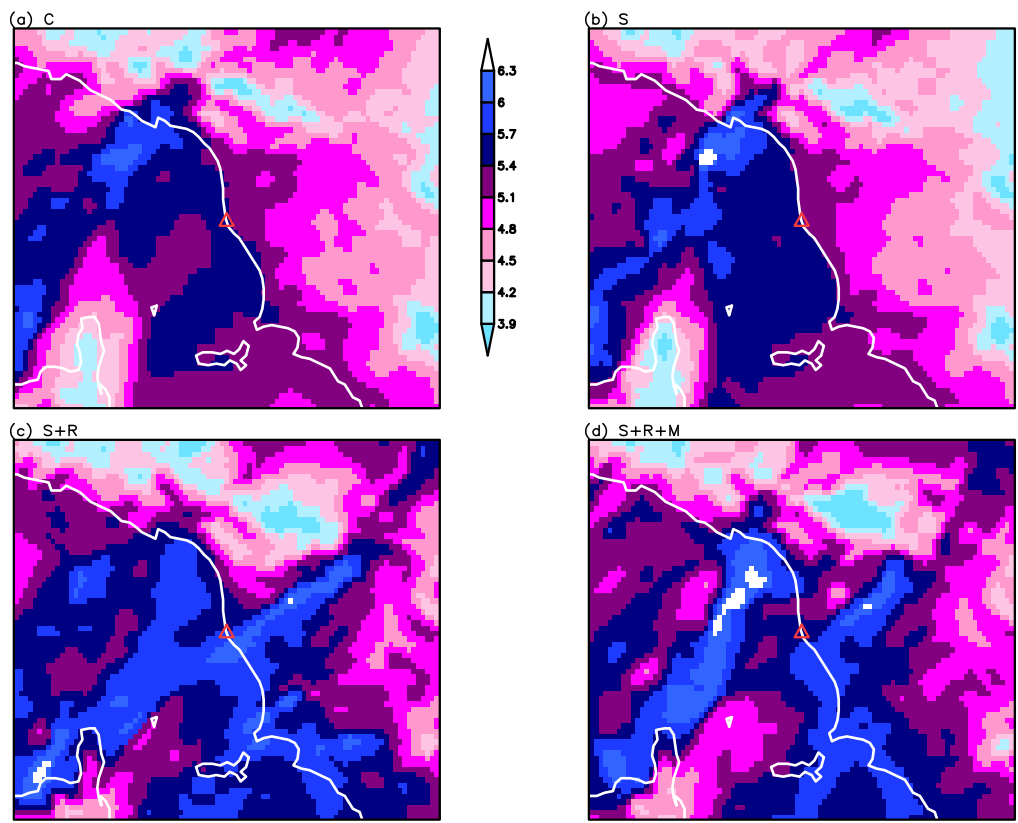}
\caption{Mean specific humidity (unit g$\cdot$ kg$^{-1}$) over the entire vertical profile at 0000 UTC of the 10th of September for (\textbf{a}): control run $C$, (\textbf{b}): assimilated run $S$, (\textbf{c}): assimilated run $S+R$, (\textbf{d}): assimilated run $S+R+M$. The red triangle (``\textcolor{red}{$\bigtriangleup$}'') indicates the Livorno town.}
\label{fig:diff:sh}
\end{figure}

To understand how the precipitation field is modified in the experimental runs, in Figure \ref{fig:diff:sh}, we show the specific humidity averaged over the entire column of atmosphere at 0000 UTC on the 10th of September. We stress the fact that these maps are the results of both the assimilation procedure and the model evolution in the first hours of integration. The $S+R$ run (panel (c) in Figure \ref{fig:diff:sh}) clearly shows a tongue of high water vapor intensity extending from the northern tip of the Corsica Island to the Tuscany northern coasts (and further inland), which is not simulated nor in the control run $C$ (panel (a) in Figure \ref{fig:diff:sh}) and in the assimilated run $S$ (panel (b) in Figure \ref{fig:diff:sh}). The addition of the simplified marine model to the $S+R$ experiment (panel (d) in Figure \ref{fig:diff:sh}) moves this line of high water vapour content northwards, close to the Ligurian coasts, slightly increasing its intensity, with values greater than 6.3 g$\cdot$ kg$^{-1}$.
%
\section{Discussion}\label{sec:discuss}
As regards the infamous rainfall event occurred in the Livorno area (Central Italy) during the night between 9 and 10 September 2017, we evaluated possible improvements arising from the assimilation of both conventional and radar data in the  setup of the WRF model, operational at the meteorological agency of the Tuscany region (LaMMA) at the time of the event. In addition, we implemented a simplified ocean model, having a relatively low computational cost, to evaluate any further improvements due to a better description of energy exchanges between air and sea, which are known to play a crucial role in the formation and intensification of HPEs \cite{foussard2019response,chelton2004satellite,o2003observations}. Such experimental forecasts were compared against a control run, which didn't assimilate any data neither implement the simplified ocean model.

Since no reliable gridded observed rainfall dataset was available for the Livorno case, as ground-truth data, we used the observations collected at the rain-gauges managed by the Hydrological Service of Tuscany \cite{capecchi2012fractal}. To assess the accuracy of QPFs, we introduced a novel method (see the sketch in Figure \ref{fig:pacman_method}), aimed at a joint evaluation of both the intensity and position errors of the forecasts, while preserving the recorded rainfall data (i.e., no interpolation or filtering of the observations was performed to comply with the model data). To further validate the predictions, we also computed standard verification scores and constructed performance diagrams for predefined precipitation thresholds.

Figure \ref{pacman}a demonstrates that the control simulation $C$ is not able to correctly predict rainfall peaks in the Livorno area; this is not new and confirms the findings of similar studies  \cite{ricciardelli2018analysis,federico2019impact,lagasio2019predictive}. However, the results shown in Figure \ref{pacman} and Table \ref{tab:rottra} suggest that the $S+R$ forecast is the more valuable. In fact, looking at the position error, the $d_0$ parameter is the lowest, indicating the optimal positioning of the transformed black circle close to the original gray circle. We note that the parameter $d_0$ for the $S+R$ forecast is about one third the length of $d_0$ for the $S$ forecast and about one quarter the length of $d_0$ for the $C$ control run, indicating the relatively higher accuracy of $S+R$ with respect to $C$ and $S$. For all the forecasts, with the exception of the $S+R$ case, the $RMSE$ minimisation transformation moves the circle including the representative rain observation points inland, next to the area with the highest predicted rain and applying a translation with a slight rotation (the parameter $\phi$ is close to $360^\circ$). In the $R+S$ case, on the contrary, the translation moves the circle towards the seaside with a rotation of $\phi=196^\circ$ that flips coastline and inland observation points, reversing North with South. Indeed the $S+R$ forecast locates the heaviest rain inland, instead of near the coast as actually happened and  further South. Consequently the minimisation process tends to rotate the stations circle by about 180$^\circ$, flipping coastline and inland rain-gauges, as well as northern with southern ones. The translation towards seaside indicates that the heaviest rain predicted inland is even heavier than the maximum values recorded on the coast. As regards the intensity error, the lowest {$RMSE_{min}$} value is obtained for the $S$ forecast; however all the forecasts provide comparable intensity errors (at least for the selected rain-gauges).

Similar conclusions can be drawn looking at the results shown in Table \ref{tab:verif} and Figure \ref{fig:perfdiag}. Both the $S+R$ and $S+R+M$ forecasts outperform $C$ and $S$, yielding better scores. However, in lack of any assessment regarding the position errors, it is difficult to evaluate which forecast between the $S+R$ and $S+R+M$ forecast provide the more valuable predictions. We deduce that the position/intensity error information shown in Table \ref{tab:rottra} and Figure \ref{pacman} confirm and complement the information obtained from standard verification skill scores shown.

Our conclusions agree with those of \cite{federico2019impact}, who found that assimilating radar (and lighting) data provide better QPFs by changing the water vapour mixing ratio during the assimilation step. Incidentally, we note that the performance diagram shown in Figure \ref{fig:perfdiag}b well agrees with Figure 16f in \cite{federico2019impact} and provides similar, or slightly better, results, supporting common conclusions. However, we acknowledge that a direct comparison should be treated with care, because ingested radar data and assimilation techniques are different and because the period over which rainfall data are accumulated is shorter (3 vs. 6 hours).

Our findings further confirms \cite{xiao2007multiple,maiello2014impact,tian2017assimilation,federico2019impact,lagasio2019predictive} that 3D-Var radar data assimilation is an effective method for improving QPFs by correcting the initial conditions of limited-area weather numerical models. In fact, concerning the assimilation of radial velocity data (available at the Aleria S-band radar only), Figure \ref{figure_wind} suggests that the impact of remote sensed data is beneficial for a better positioning of the low-level winds convergence line. Moreover, the indirect assimilation of radar reflectivity (available at both the Aleria and Livorno radars) and in turn the modification of water vapour mixing ratio (see Figure \ref{fig:diff:sh}) is supposed to provide an environment that is conducive for convection; this latter carries towards higher atmospheric levels large contents of moisture (see panels (c) and (d) in Figure \ref{fig:diff:sh}) and energy favoring cloud formation. Thus, water vapour spatial and temporal variation indirectly provides some indications on how the different model setups may impact the event prediction. Our findings agree with those of \cite{xiao2007multiple}, who discussed how radial velocity assimilation has a large impact on wind velocity, whereas reflectivity data has a direct impact on hydrometeor analyses. We conclude that, since the origin of the Livorno event took place on the sea, the assimilation of conventional data, available only on land, has a low impact on the accuracy of QPF for the $S$ experiment. On the other hand, assimilating radar data, which are available offshore, provides pertinent and crucial information regarding rainfall triggering and atmospheric moisture content. Furthermore, we claim that such remote sensed data modifies the model dynamics in a way that persists few hours (up to 9 hours) after the assimilation step (which ends at 1800 UTC on the 9th of September).

When looking at the outputs of the $S+R+M$ experiment (see panels (d) in Figures \ref{pacman}, \ref{figure_wind} and \ref{fig:diff:sh}), the forecasts are not dramatically affected by the use of the simplified ocean model. This is consistent with the conclusions drawn by \cite{lebeaupin2006sensitivity}, who find that the evolution of the SSTs during the model integration has a marginal impact on the short-range predictions (forecast length less than 18 hours). We can deduce that, for the Livorno case, the exchange of both heat and water vapour between air and sea doesn't play a crucial role as also found by  \cite{lagasio2019synergistic}. In fact, the authors argued that ingesting SSTs estimates from Sentinel-3 satellite observations into the WRF model do not improve the Livorno QPF for high precipitation rate. This happens possibly because at the kilometre scale, the SSTs determine the intensity of the warm low-level jet \cite{ricchi2016use}, which, in our case, was partially corrected by the ingestion of radial velocity radar data. In our $S+R+M$ forecast, it is observed that the heat southerly fluxes are more intense by about 10-15\% if compared to $S+R$ data (map not shown). We found that, although the skill scores of $S+R$ and $S+R+M$ are similar (see Table \ref{tab:verif} and Figure \ref{fig:perfdiag}), the simplified ocean model is able to modify the precipitation field by reducing the QPF peaks close to the coast and thus producing a position error (see panels (c) and (d) in Figure \ref{pacman} and Table \ref{tab:rottra}).

We claim that assimilating radar data turned out to be effective because we used a covariance matrix of the background errors $\mathbf{B}$, that is the result of a long-term application of the NMC method (approximately ten months). In fact, to estimate $\mathbf{B}$, we used the operational forecasts issued twice a day at the regional meteorological service of Tuscany (LaMMA), which share the same setup (in terms of resolution, number of vertical levels, physical parameterisations, etc\ldots) of the runs here presented. This is a remarkable improvement with respect to previous similar studies; to mention a few works \cite{maiello2014impact} and  \cite{lagasio2019predictive} employed a 1-week and a 1-month period, respectively, to compute the $\mathbf{B}$ matrix, \cite{federico2019impact} and \cite{mazzarella2020reflectivity} applied the NMC method during the Hydrological cycle in the Mediterranean Experiment – First Special Observing Period (HyMeX-SOP1, which lasted for about two months in 2012, see \cite{ducrocq2014hymex} ), whereas \cite{tian2017assimilation} used the default matrix provided by the WRFDA system, which is produced with global data and its use for regional cases is sometimes discouraged \cite{skamarock2019description}. Also more recent works (see  \cite{mazzarella2021investigating}) used a time period shorter than one month to compute the covariance matrix of the background errors. However, we acknowledge that further investigations are needed to assess the impact of a long-term application of the NMC method to build a high-level quality $\mathbf{B}$ matrix and obtain better QPF data.
%
\section{Conclusions}\label{sec:conc}
Although tremendous improvements were achieved in recent years, the operational forecasting of HPEs in the Western Mediterranean basin still remains a challenging task, in particular in the context of a warming climate. As concerns the Livorno case, the main findings of this work are:
\begin{itemize}
    \item assimilating reflectivity data from X- and S-band radars and radial velocity data from S-band radar significantly improves the description of atmospheric humidity content and low-level winds, resulting in better QPFs,
    \item the application of a simplified ocean model, although modifies the low-level jet associated with the event, scarcely impacts the short-range forecast (length shorter than 12 hours) of precipitation,
    \item the novel QPF verification method introduced in this paper, based on roto-translation  RMSE-minimisation, confirms and reinforces the results achieved with standard verification scores, adding more information about the position error of the WRF simulations.
\end{itemize}
The conclusions we drew support the deployment of a dense network of relatively small radars in the coastal areas of the Western Mediterranean Sea. Such high-resolution (both spatial and temporal) data may have beneficial impacts on short-range numerical weather predictions. In fact, by extracting the maximum value from local observations, such as those collected by X-band radars which are not processed by international meteorological organizations, higher quality analyses can improve the description of finer scale features and thus provide better initial conditions to limited-area weather models.

\vspace{6pt}




\acknowledgments{The Copernicus Climate Change Service is acknowledged for the ERA5 data, which were used to produce the maps in Figure \ref{figure_one}. This study has been conducted using E.U. Copernicus Marine Service Information. Rain-gauge data were provided by the meteorological network managed by the Hydrological Service of Tuscany. 
Antonio Ricchi funding has been provided via the PON ``AIM'' - Attraction and International Mobility program AIM1858058. }


\reftitle{References}
\externalbibliography{yes}
\bibliography{liv2017_water.bib}

\end{document}